\documentclass{aa}
\pdfoutput=1
\usepackage{txfonts,textcomp}
\usepackage{graphicx}
\usepackage{color}
\usepackage{tikz}
\usetikzlibrary{shapes,arrows}
\usepackage{multirow}
\usepackage{natbib}
\bibpunct{(}{)}{;}{a}{}{,}

\begin{document}
\title{Detecting non-uniform period spacings in the \emph{Kepler} photometry of $\gamma$ Doradus stars: methodology and case studies\thanks{Based on data gathered with the NASA Discovery mission \emph{Kepler} and the HERMES spectrograph, which is installed at the Mercator Telescope, operated on the island of La Palma by the Flemish Community at the Spanish Observatorio del Roque de los Muchachos of the Instituto de Astrof\'isica de Canarias, and supported by the Fund for Scientific Research of Flanders (FWO), Belgium, the Research Council of KU Leuven, Belgium, the Fonds National de la Recherche Scientifique (F.R.S.-FNRS), Belgium, the Royal Observatory of Belgium, the Observatoire de Gen\`eve, Switzerland, and the Th\"uringer Landessternwarte Tautenburg, Germany.}\fnmsep\thanks{Table A.1 is only available in electronic form at the CDS via anonymous ftp to cdsarc.u-strasbg.fr (130.79.128.5) or via http://cdsweb.u-strasbg.fr/cgi-bin/qcat?J/A+A/}}

\author{T.~Van~Reeth\inst{1}
\and A.~Tkachenko\inst{1}\thanks{Postdoctoral Fellow of the Fund for Scientific Research (FWO), Flanders, Belgium.}
\and C.~Aerts\inst{1,2}
\and P.~I.~P\'apics\inst{1}\thanks{Postdoctoral Fellow of the Fund for Scientific Research (FWO), Flanders, Belgium.}
\and P.~Degroote\inst{1}\thanks{Postdoctoral Fellow of the Fund for Scientific Research (FWO), Flanders, Belgium.}
\and J.~Debosscher\inst{1}
\and K.~Zwintz\inst{1}
\and S.~Bloemen\inst{2}
\and K.~De~Smedt\inst{1}
\and M.~Hrudkova\inst{3}
\and G. Raskin\inst{1}
\and H. Van Winckel\inst{1}}
\institute{Instituut voor Sterrenkunde, KU Leuven, Celestijnenlaan 200D, 3001 Leuven, Belgium
\and Department of Astrophysics, IMAPP, University of Nijmegen, PO Box 9010, 6500 GL Nijmegen, The Netherlands
 \and Isaac Newton Group of Telescopes, Apartado de Correos 321, 387 00 Santa Cruz de la Palma, Canary Islands, Spain}
\authorrunning{T. Van Reeth et al.}
\titlerunning{Detecting non-uniform period spacings in $\gamma$ Dor stars}

\date{Received 11 July 2014 / Accepted 24 October 2014}

\abstract{ The analysis of stellar oscillations is one of the most reliable
  ways to probe stellar interiors. Recent space missions such as \emph{Kepler}
  have provided us with an opportunity to study these oscillations with
  unprecedented detail. For many multi-periodic pulsators such as $\gamma$
  Doradus stars, this led to the detection of dozens to hundreds of oscillation
  frequencies that could not be found from ground-based observations.}
{We aim to detect non-uniform period spacings in the Fourier spectra of a sample
  of $\gamma$ Doradus stars observed by \emph{Kepler}. Such detection is
    complicated by both the large number of significant frequencies in the space
    photometry and by overlapping non-equidistant rotationally split
    multiplets.}
  {Guided by theoretical properties of gravity-mode oscillation of $\gamma$
    Doradus stars, we developed a period-spacing detection method and applied it
    to {\it Kepler\/} observations of a few stars, after having tested the
    performance from simulations.}
  {The application of the technique resulted in the clear detection of non-uniform
    period spacing series for three out of the five treated {\it Kepler\/} targets.
      Disadvantages of the technique are also discussed, and include the
      disability to distinguish between different values of the spherical degree
      and azimuthal order of the oscillation modes without additional
      theoretical modelling.}
{Despite the shortcomings, the method is shown to allow solid
    detections of period spacings for $\gamma$ Doradus stars, which will allow
  future asteroseismic analyses of these stars.}


\maketitle

\section{Introduction}
\label{sec:intro}
Gamma Doradus pulsators are early F- to late A-type main-sequence stars. This
places them directly within the transition region between low-mass stars with a
fully convective envelope and higher-mass stars with a convective core. They
exhibit non-radial gravity modes which are excited by the flux blocking
mechanism at the base of their convective envelopes
\citep[e.g.][]{Guzik2000,Dupret2005}. The oscillations penetrate
throughout the radiative zone between the stars' convective core and
envelope. The pulsation periods typically vary between 0.3 and 3 days
  \citep[e.g.][]{Kaye1999}, though recently it has been predicted by theory
that they can be shifted outside of this range because of rotation
\citep{Bouabid2013}.

While various theoretical studies have been conducted on the internal structure
of A- to F-type main-sequence stars, there have been few possibilities to verify
the theoretical models observationally \citep[e.g.][and references
therein]{Miglio2008a}. Gamma Doradus pulsators have potential in this
  respect, since modelling of their oscillations would enable to probe their
internal structure \citep{Miglio2008a,Bouabid2013}. Such successful
  modelling based on g-mode period spacings detected in space photometry has
  been accomplished for main-sequence B-type stars
  \citep[e.g.][]{Degroote2010,Savonije2013,Papics2014}, sdB stars
  \citep[e.g.][]{Reed2011}, subgiants and red giants
  \citep[e.g.][]{Beck2011,Beck2012,Bedding2011,Mosser2012,Deheuvels2014},
thanks to 
data from the space missions  CoRoT \citep{Auvergne2009} and
  \emph{Kepler} \citep{Koch2010}. Period spacings have been reported for
  very few AF-type main-sequence pulsators so far, e.g.
  \citet{Chapellier2012,Kurtz2014}, while some detailed searches led to a
  failure to detect this important diagnostic \citep{Hareter2012,Breger2012} 
  because of too poor frequency resolution of the data or too rapid rotation 
  of the star.

For $n\gg l$, with $n$ the radial order and $l$ the spherical degree of the
  mode, the first-order asymptotic approximation predicts the gravity modes
    to be equally spaced in period \citep{Tassoul1980}. Chemical gradients
    induced by the changing size of the stellar core during the stars' evolution
    \citep{Miglio2008a}, as well as rotation and mixing processes
    \citep{Bouabid2013} complicate this simple picture and lead to
  characteristic deviations from uniform period spacings.

In this paper, we present a method for detecting non-uniform period spacings in the
frequency spectra of $\gamma$ Doradus pulsators (Section \ref{sec:method}),
guided by some general theoretical expectations and previous observational
studies of such stars (Section \ref{sec:theorexp}).  We test the method on
simulated light curves and apply it to \emph{Kepler} data of five $\gamma$
Doradus stars in Section \ref{sec:app}. Finally, we summarise the methodology
and give our conclusions and future prospects.

\section{Theoretical expectations}
\label{sec:theorexp}
In the first-order asymptotic approximation of a non-rotating star, the 
periods of high-order ($n\gg l$) g-modes in stars with a convective core and a
radiative envelope \citep{Tassoul1980} are defined as 
\begin{equation}
P =
\frac{\Pi_0}{\sqrt{l\left(l+1\right)}}\left(n + \alpha_{l,g}\right),
\end{equation}
 with
\begin{equation}
\Pi_0
= 2\pi^2\left(\int_{r_1}^{r_2}N\frac{\mathrm{d}r}{r}\right)^{-1}.
\label{eq1}
\end{equation}
 Here
$\alpha_{l,g}$ depends on the boundaries of the trapping region, $P$ is the
oscillation period, $r$ the distance from the stellar centre, and $N$ the
Brunt-V\"ais\"al\"a frequency.
Following this relation, 
modes with the same degree $l$, but consecutive orders $n$, are
equidistantly spaced in period. Their spacing value $\Delta \Pi$ is then given
by 
\begin{equation}
\Delta \Pi_l = \frac{\Pi_0}{\sqrt{l\left(l+1\right)}}.
\label{eq2}
\end{equation} 
As a result, the
expected period spacings for different $l$-values will have fixed ratios.

\begin{figure}
 \includegraphics[width=88mm]{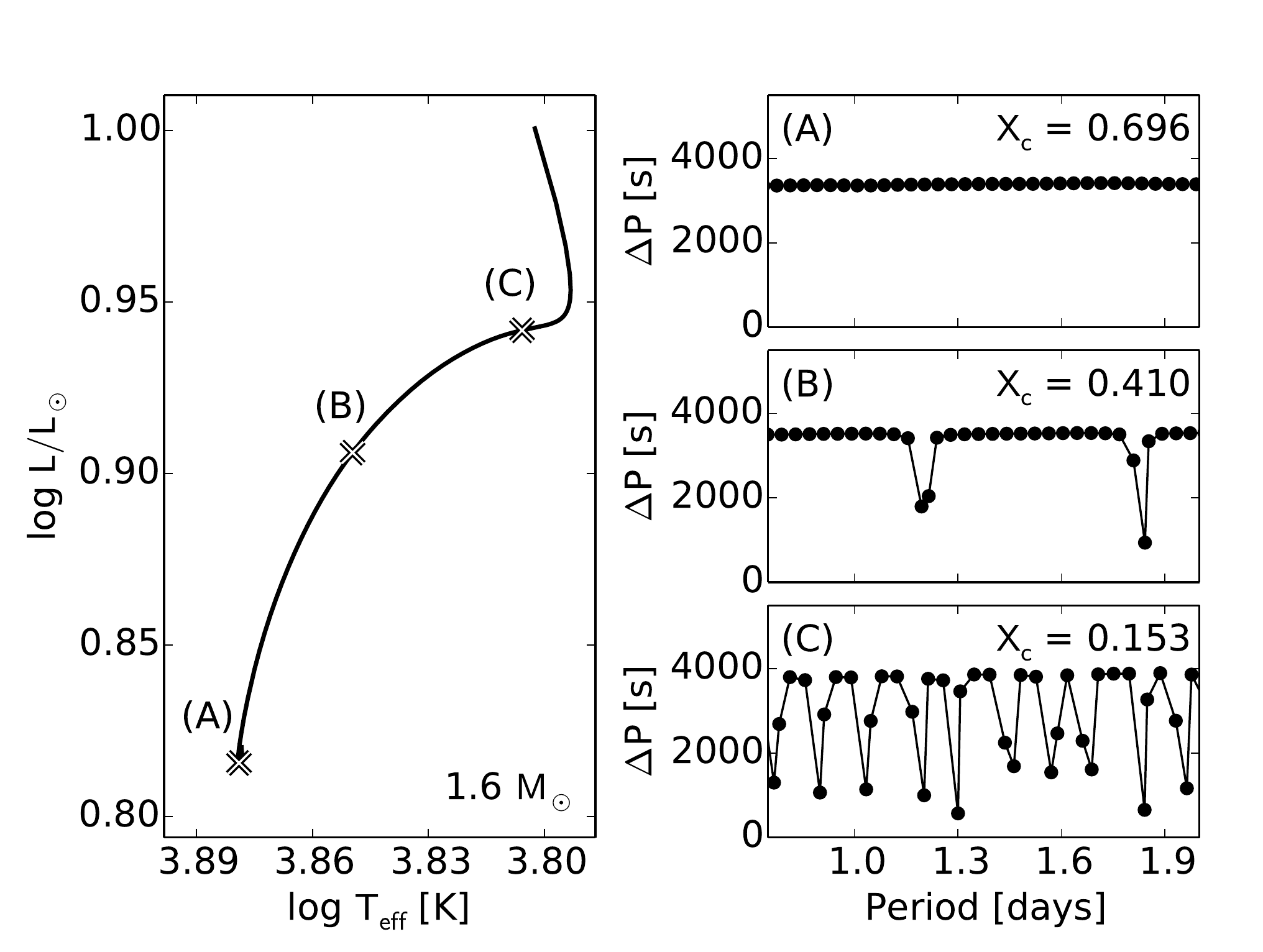}
 \caption{\label{fig:miglio}\textbf{Left:} Part of a 1.6\,$\rm M_{\odot}$ evolution track, computed with MESA. \textbf{Right:} Period spacing patterns computed for dipole modes of the marked models on the evolution track in the left plot. For each the hydrogen content of the core ($X_c$) is provided.}
\end{figure}

Several aspects,  which are not treated in this simple approximation, influence 
the period spacings, however. For instance, during the star's core-hydrogen 
burning phase, the convective core can either grow or shrink, depending on the 
star's birth mass \citep[Fig.3.6 in][]{Aerts2010}. A growing convective core 
leads to a discontinuity in the chemical composition at its boundary, while a 
receding convective core leaves behind a $\mu$-gradient zone near the core. 
\citet{Miglio2008a} studied the impact of such a chemical gradient on the 
period spacings, discussing also in detail the way to define the convective 
boundary, which we therefore do not repeat here.
The authors showed that the variation of the local average molecular weight 
leads to modifications of the g-mode resonance cavity. This mode trapping
translates into characteristic dips in the period spacing pattern. The
amplitude of the dips indicates the steepness of the chemical gradient, whereas
the periodicity of the dips indicates its location. This is illustrated in
Fig. \ref{fig:miglio}, where we show the period spacing patterns calculated with
the pulsation code GYRE \citep{Townsend2013} for a 1.6\,$\rm M_\odot$ star at
different phases of its evolution. The models were computed with MESA
version 6208 \citep{Paxton2011,Paxton2013}. We used the standard MESA input
physics with the Schwarzschild criterion for convection and an overshoot value
of 0.014 local pressure scale heights in the formulation by 
\citet{Herwig2000}.

\begin{figure}
 \includegraphics[width=88mm]{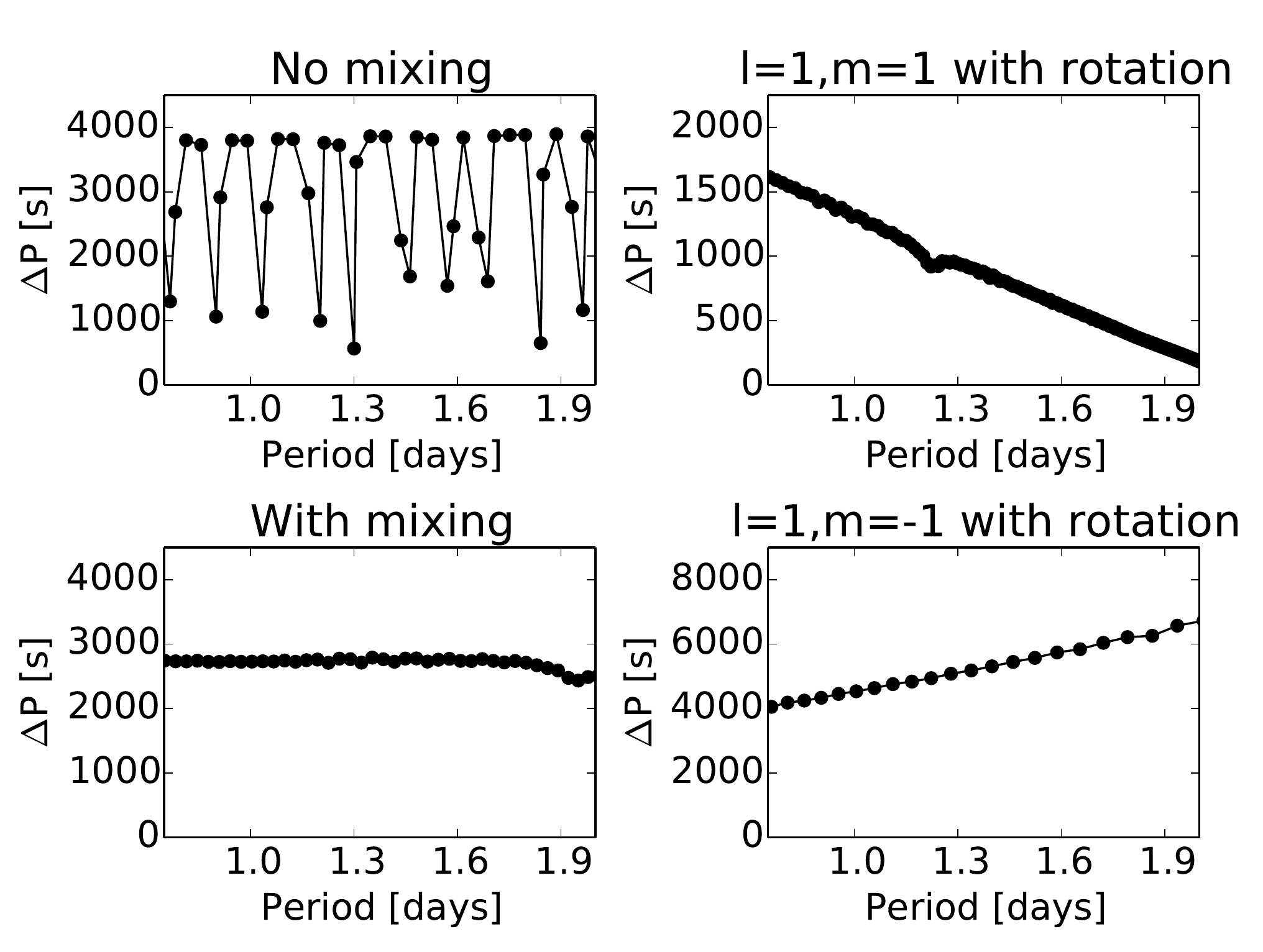}
 \caption{\label{fig:bouabid}Period spacing patterns computed for dipole modes
   of a 1.6\,$\rm M_{\odot}$ star with $X_c = 0.153$. \textbf{Left:} spacing
   patterns with and without diffusive mixing. \textbf{Right:} spacing patterns
   including the effect of diffusive mixing for a star rotating with $v_{\rm eq} =
   73$\,km\,$\rm s^{-1}$ for the prograde (top) and retrograde (bottom)
   modes. We note the difference of the scales on the y-axes.}
\end{figure}

\citet{Bouabid2013} further extended the work of \citet{Miglio2008a}, and
included mixing processes and rotation. The internal mixing processes wash out
the chemical gradient, which reduces the presence of the dips in the spacing
pattern and decreases the average period spacing value, though there
  might still be dips present in the period spacing pattern.  The influence
  of rotation varies from mode to mode.  The period spacings of prograde modes,
  which travel in the direction of rotation, will become smaller because of the
  positive rotational frequency shift added to the oscillation frequencies in an
  inertial frame.  Retrograde modes, which travel in the opposite direction of
  rotation, will have larger period spacings than zonal modes. This is
illustrated in Fig. \ref{fig:bouabid} for dipole modes. Diffusive mixing was 
taken into account according to standard MESA description 
\citep{Paxton2011,Paxton2013}, while the frequency shifts resulting from rotation were 
computed using the approximation provided by \citet{Chlebowski1978}, which is 
too simplistic to model the frequencies of a real star in full detail but 
sufficiently appropriate to illustrate the effects of rotation on prograde 
and retrograde modes as is our purpose here.

So far most of the mode identification studies of $\gamma$ Doradus stars
  relied on spectroscopic time series \citep[e.g.][]{Brunsden2012,Davie2014}.
By far the most frequently found identification are progade dipole modes.  We
therefore expect to find period spacing patterns as described by
\citet{Miglio2008a} and \citet{Bouabid2013}, corresponding to prograde dipole
modes. These theoretical and spectroscopic results are used as a guide to
  develop a new method for the detection of period spacings from uninterrupted
  high-precision space photometry.

\section{The methodology}
\label{sec:method} 
\begin{figure}
 \includegraphics[width=88mm]{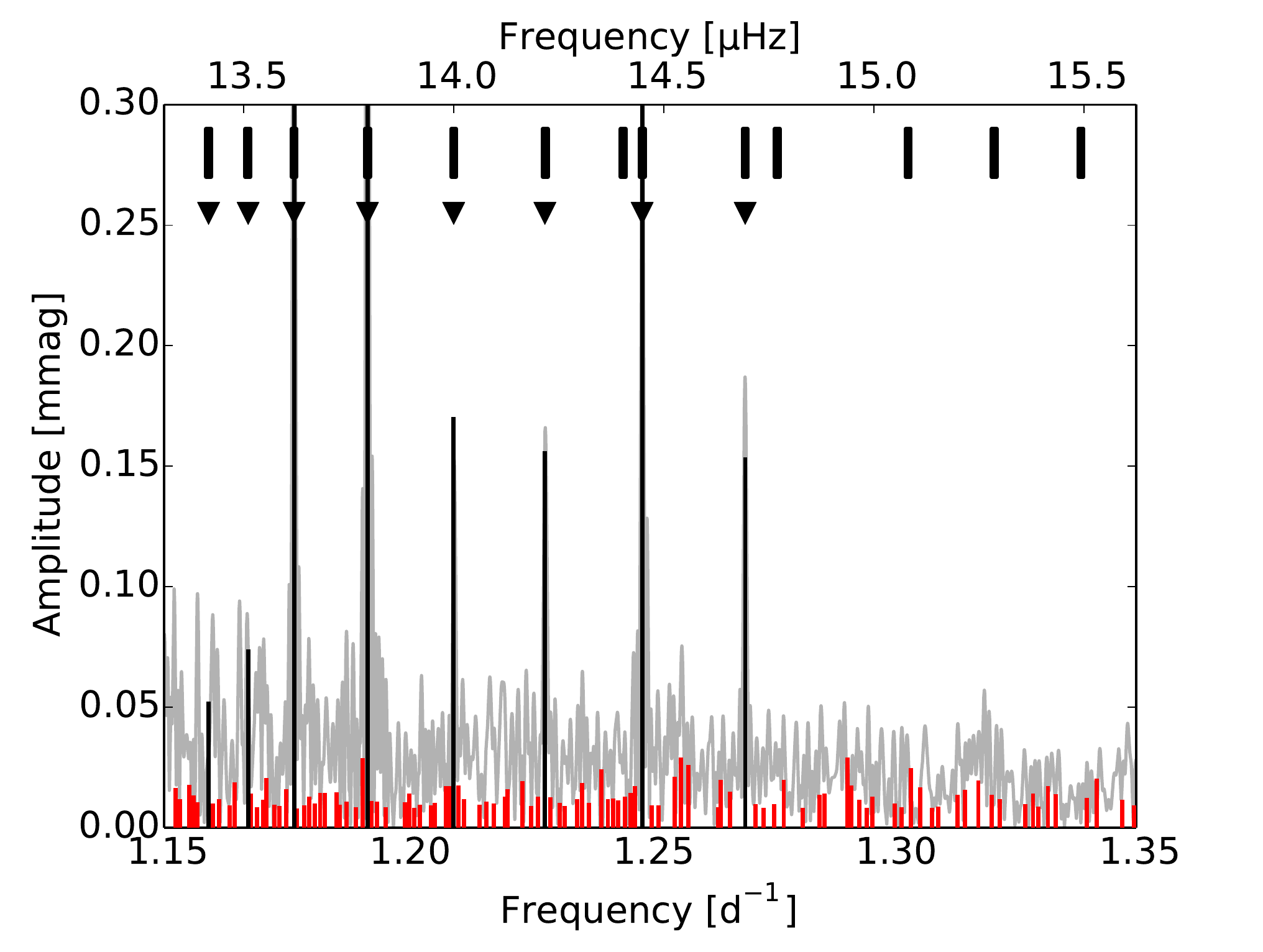}
 \caption{\label{fig:tradprew}Part of the Fourier transform (grey) of a
   simulated light curve. The full black vertical lines indicate the frequencies
   which were extracted using traditional prewhitening with S/N\,$\geq$\,8.0,
   while the red vertical lines indicate the remaining frequencies which were
   extracted with S/N\,$\geq$\,4.0. The thick black markers indicate the
   frequencies which were used as input, and the black triangles show which
     frequencies were extracted with our definition of the comparison criterion.}
\end{figure}

While it has long been known that $\gamma$ Doradus stars
  are multi-periodic pulsators with at least {\it a few\/} modes
  \citep[e.g.][]{Handler1999,Kaye1999,Cuypers2009}, space-based
  observations revealed dozens to hundreds of oscillation frequencies for each
  star. This in principle should allow the detection of period spacing
  patterns. The high precision of the data implies that the prewhitening method
  to accept frequencies with an amplitude up to four times the local noise level
  in the Fourier transform, as originally suggested to detect high-frequency
  p-mode pulsations from
  ground-based photometric campaigns of $\delta\,$Sct pulsators by
  \citet{Breger1993},  is not necessarily suitable
  in the framework of space photometry data in general, and particularly not in
  the search for period spacings of g-mode pulsators 
\citep[e.g.][]{Balona2014}.  It would lead to
  the extraction of frequency values that are seriously influenced by the
  preceding prewhitening. This is illustrated with simulations mimicking a
  typical {\it
    Kepler\/} light curve of a $\gamma\,$Doradus star in our sample 
in Fig. \ref{fig:tradprew}. A lot of
  low-amplitude peaks are derived from the power spectrum of  the 
simulated light curve and its subsequent prewhitening,
while they are unrelated to the simulated
  signal.  For this reason, recent analyses of {\it Kepler\/} photometry 
either do not use a stop criterion for the frequency analysis
  \citep[e.g.][]{Kurtz2014}, or rely on a 
higher cutoff value than four times the local noise level. 
In Fig. \ref{fig:tradprew}, we show all frequencies whose amplitudes have a S/N
level above eight.

The suitable S/N cutoff value varies strongly from star to star, and is not
  known a priori. The reason is that for such high-quality data, noise is no
  longer the limiting factor for the accuracy of the result. Rather, small
  uncertainties in the frequency, amplitude and phase values when fitting sine
  functions to the light curve give rise to a residual signal. These residuals
  modulate the subsequently extracted frequency values and their influence keeps
  increasing during the iterative prewhitening procedure.  This is our
  motivation to deduce a new stop criterion based on the comparison of the
  extracted frequencies with the original Fourier spectrum of the light
  curve. Hereafter, we call this the comparison criterion.

\begin{figure}
 \includegraphics[width=88mm]{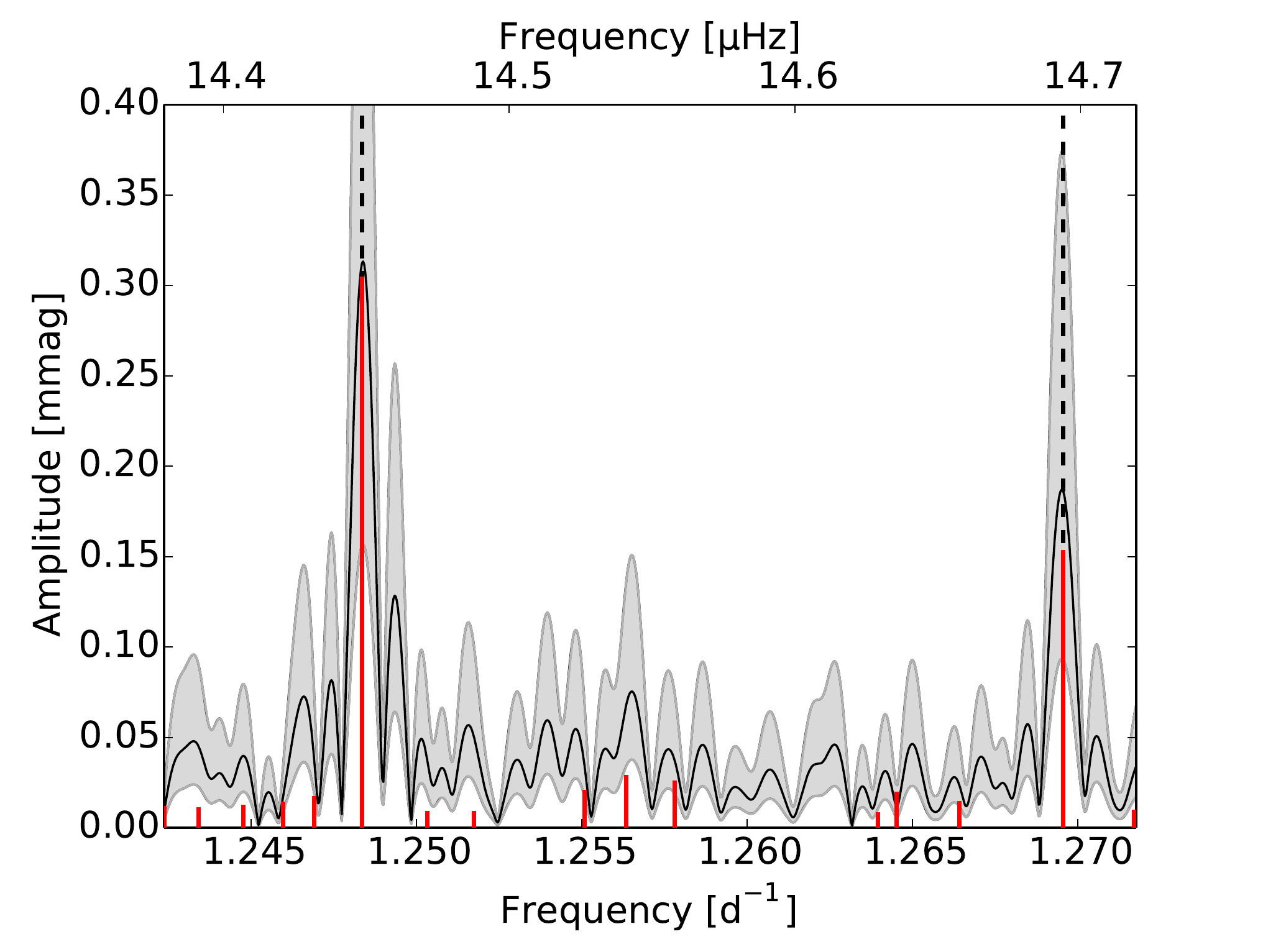}
 \caption{\label{fig:stopcrit}The Fourier transform (full black curve) of a simulated light curve. Using the presented comparison criterion, the frequency extraction is aborted when the amplitude of an extracted frequency (full red vertical lines) does not agree with the original Fourier transform within a certain margin (grey area). The accepted frequencies are marked with dashed lines.}
\end{figure}

We first compute the amplitude $A_f$ of the
extracted frequency $f$ by fitting a sine function with that frequency to the
light curve. Then we compare $A_f$ to the local amplitude $A_{loc}$ of the
Fourier spectrum for the frequency $f$, and only accept the extracted frequency
if both amplitudes agree within a certain margin (see
Fig. \ref{fig:stopcrit}): 
\begin{equation}
\alpha \leq \frac{A_f}{A_{loc}} \leq
\frac{1}{\alpha}.
\label{stop}
\end{equation}
 Here $\alpha$ is given a reasonable value, which we found to
be 0.5 for most stars. When the considered frequency peak does not fulfill this
requirement, the frequency extraction is aborted. While it is likely that, with
this criterion, the residuals of the light curve will still contain a lot of
signal, the modulation of this signal by the preceding prewhitening may no
longer be negligible. By taking alpha = 0.5, we are able to keep the number 
of detected false frequencies down to a minimum. For most 
$\gamma$\,Doradus stars the extracted frequencies
are a good starting point for modelling, in the sense that 
they are sufficient to look for part of the period
spacing series expected to be present in the data.

Detecting non-uniform period spacings can be rather complicated, especially if
two different series with a different average spacing overlap. However, there
are several aspects which may facilitate the detection. It is for example
well-known that the visibility of a stellar oscillation mode depends on its
degree $l$ and order $m$, and on the inclination angle and rotational velocity
of the star \citep[e.g.][]{Chadid2001}. As a result, many of the visible
oscillation modes are likely to have similar values for $l$ and $m$ while
differing in radial order. As we
already noted in Section \ref{sec:theorexp}, we expect prograde sectoral modes
with low $l$ values to be the most easily visible \citep{Townsend2003}. In
addition, early F- to late A-type main-sequence stars are typically moderate to
fast rotators. While this means that the rotational splitting of oscillation
frequencies will be asymmetric and thus difficult to detect (especially
considering the high density of the frequency spectra), it also entails that
when the star is sufficiently fast rotating, its frequency splitting will 
be larger than the frequency region where the pulsation frequencies of 
similar slowly rotating pulsators would be located. As a 
result, we can detect groups of frequencies within the frequency spectra, 
each of which containing 
mostly frequencies with the same $l$ and $m$ values.
Finally, many $\gamma$ Doradus pulsators have strongly asymmetric light curves 
\citep{Balona2011}. As a result, several of the peaks we
detect in the Fourier spectrum, correspond to combinations of oscillation
frequencies rather than frequencies of individual pulsations
\citep{Balona2012,Papics2012b}. We exclude these combination
frequencies from the period spacing search.

One of the most efficient ways to detect period spacings for a typical
$\gamma$\,Doradus star is to order the accepted oscillation periods and
determine the spacings between them. It is likely that at least some adjacent
accepted frequency peaks have the same values for $l$ and $m$, making them part
of the same (possibly non-uniform) period spacing series. When, on the other
hand, the studied $\gamma$ Doradus star is a slow rotator, the rotational
splitting will likely be too small for this detection method to work. However,
in this case we can look for spacings and/or splittings using other traditional
methods, such as \'echelle diagrams \citep[e.g.][]{Grec1983,Mosser2013} because
the spacing pattern will be more regular. \'Echelle diagrams are not useful
  for stars rotating such that their rotational frequency shifts no longer result
  in quasi-equidistant spacing patterns. Because of the expected non-uniform nature
of the period spacings, detected spacing patterns should always be evaluated
visually by the researcher.

After looking for period spacings, we iterate over the algorithm: we go back to
the frequency extraction, but choose a different value for $\alpha$. By
iteratively choosing smaller values (e.g.\ from 0.4 to 0.05) we can extract more
frequencies and systematically look for additional frequency peaks which match
the found period spacing series. If no period spacings were detected (yet), it
is possible that choosing a smaller value for $\alpha$ will help, though one
must keep in mind that more noise peaks will be included as well, such that
probability of detecting spurious period spacings will be higher. Going
  back to Fig. \ref{fig:tradprew}, we show the frequencies extracted with this
  proposed criterion compared to the classical signal-to-noise criterion for a
  simulated light curve.  While the remaining low-amplitude input frequencies
  are also extracted with the classical criterion, the
  number of additional false frequencies is too high to allow for a reliable
  selection of the input frequencies.

\begin{figure}[h]
\includegraphics[width=88mm]{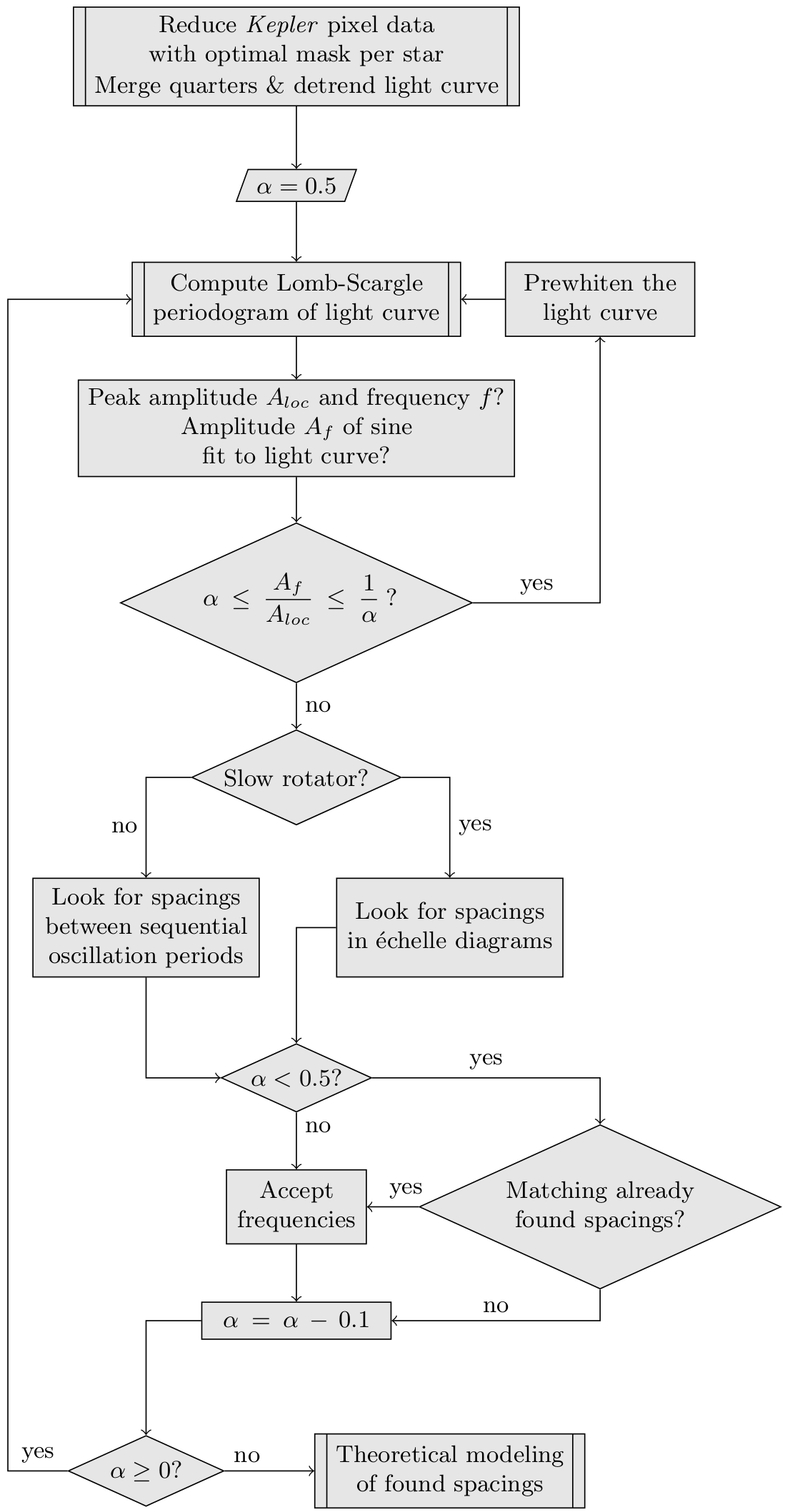}
\caption{\label{fig:methodology}Schematic summary of the methodology described in Section \ref{sec:method}. While the values of $\alpha$ provided here work well for most $\gamma$ Dor stars, they can of course be customised for individual stars.}
\end{figure}

  A summary of our method is shown schematically in
  Fig. \ref{fig:methodology}. The different steps of the method are further
  illustrated by applications on simulated data in the next Section.  The
  current version of our code does not allow us to formally identify $l$ and $m$,
  with the exception of stars for which we detect rotational multiplets. This
  can only be achieved from comparing the detected series with theoretical model
  predictions.

\section{Applications}
\label{sec:app}
\subsection{Simulations}
\label{subsec:sim}
We first evaluate our method on simulated data with two goals. 
First, we test that computing the differences between subsequent
  oscillation periods is a robust method for the detection of spacings for
  moderate to fast rotators. For slowly rotating stars, on the other hand, the
  use of \'echelle diagrams is more advantageous. Secondly, we show that the
  slope in the period spacing pattern induced by the rotation for prograde or
  retrograde modes is clearly detectable. For these tests, simulations
  simplified with respect to the 
complex physics of $\gamma$ Doradus stars are sufficient.
We therefore generated  normalised light curves
mimicking the results obtained by \citet{Miglio2008a} and
\citet{Bouabid2013}:
\begin{equation}
F(t)\ \ =\ \ A\left(\left[1 + \sum_i a_i\sin\left(2\pi
      f_it + \phi_i \right)\right]^{2.2} - 1\right) + \sigma.
\end{equation}
  Here, $f_i$ and
  $\phi_i$ are the frequency and phase of a particular oscillation mode $i$, 
  $A$ is the 
  overall amplitude of the light curve, 
  $a_i$ scales
  the amplitude of the pulsation mode $i$, and $\sigma$ represents a level of
  Gaussian white noise, similar to what is observed in the \emph{Kepler} data. The
  power 2.2 was used to simulate the asymmetry which is often seen in
  $\gamma$\,Doradus light curves \citep[see also][]{Balona2011}. 
A typical set of frequencies $f_i$ was computed from a MESA model and the GYRE 
pulsation code for both dipole and quadrupole modes taking into account the 
frequency shifts caused by rotation ($v_{eq} = 73$\,km\,$\rm s^{-1}$, $i = 77$\textdegree, 
$R = 1.7\,R_\odot$) in the approximation by \citet{Chlebowski1978}. The full list of 
freqiencies is available online in Table A.1 at the CDS. The light curve was simulated 
using the time stamps of real \emph{Kepler} observations and converted to magnitude scale.

\begin{figure}
 \includegraphics[width=88mm]{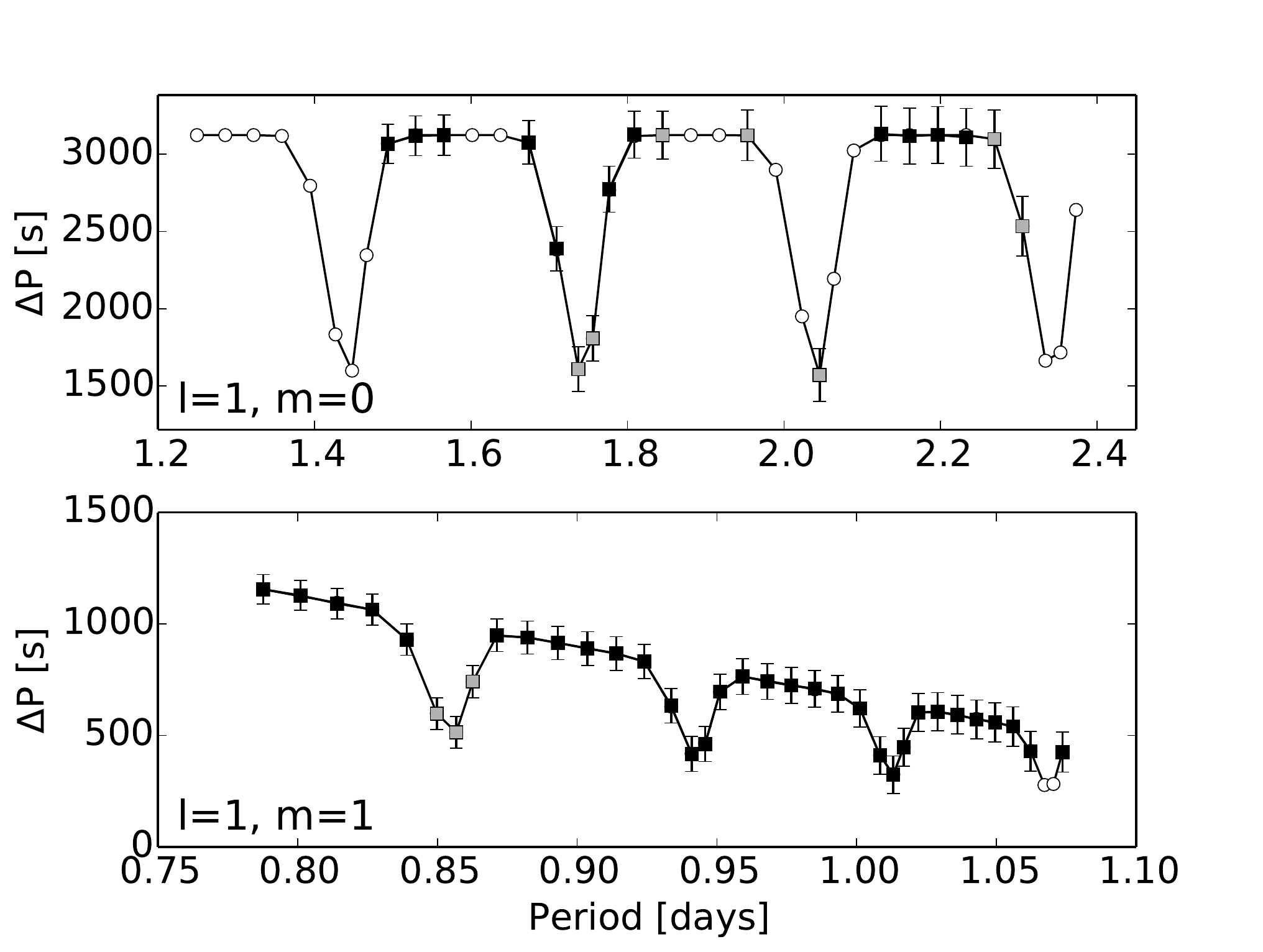}
 \caption{\label{fig:sim_sp}The period spacings corresponding to several input frequencies in a simulated data set. The found spacings with $\alpha\,=\,0.5$ are indicated by black squares, while spacings found for other values of $\alpha$ (from 0.4 to 0.1) are represented by grey squares, and non-detected spacings by white dots. This is shown for both zonal (top) and prograde (bottom) dipole modes.}
\end{figure}

\begin{figure}
 \includegraphics[width=88mm]{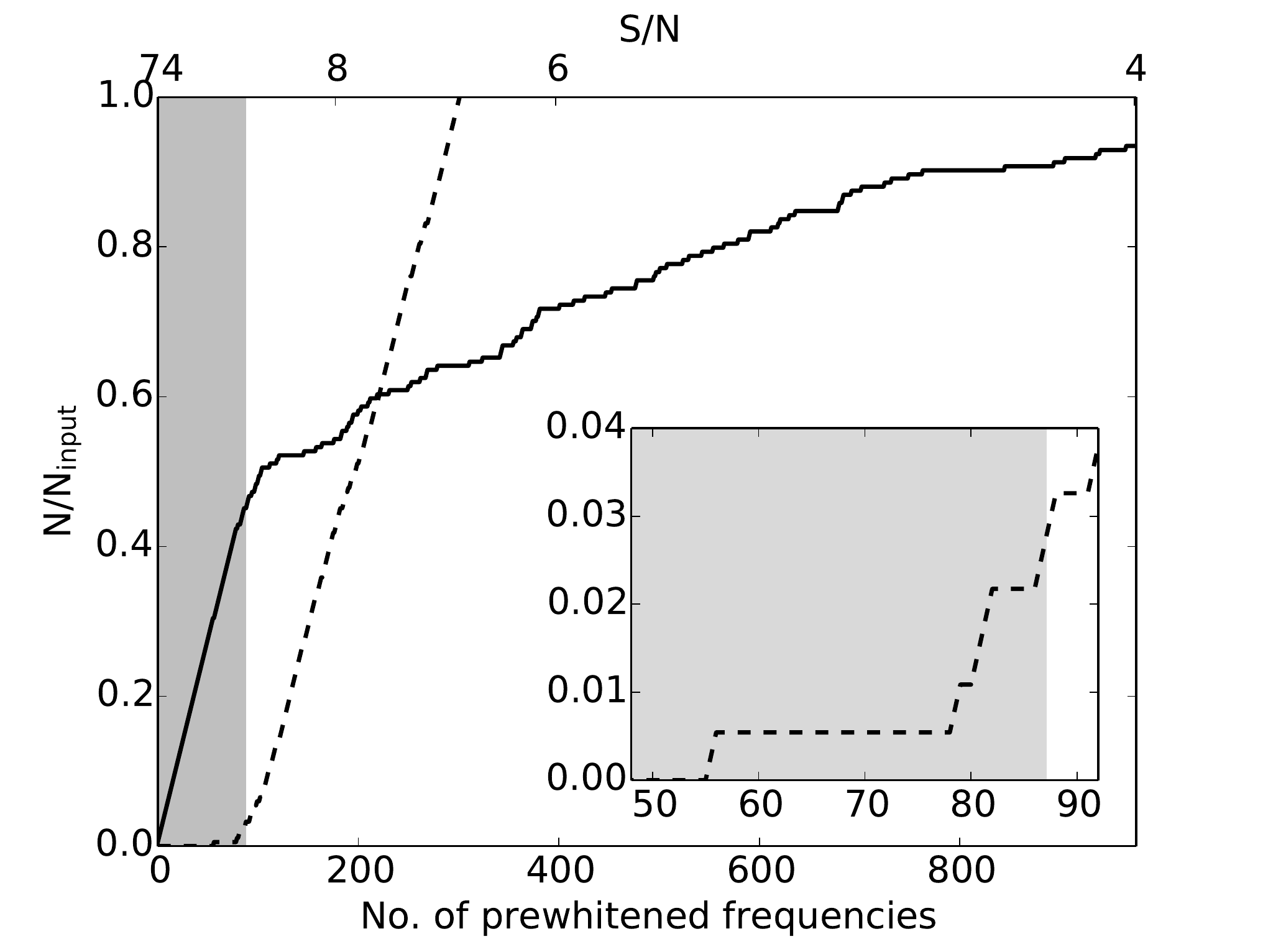}
 \caption{\label{fig:sim_comp}Fraction of extracted to input frequencies
     as a function of the number of prewhitened frequencies (full black line). The
     dashed grey line indicates the relative number of detected noise
     frequencies as a function of the number of prewhitened frequencies when
     ignoring the comparison criterion.  The upper axis provides the
     signal-to-noise ratio of the last extracted frequency. The shaded area
     indicates the number of accepted frequencies
when adopting the comparison criterion with
     $\alpha=0.5$, for which the number of extracted noise peaks remains
modest.}
\end{figure}

  We extracted the oscillation frequencies from the simulated data using a
  prewhitening method, as described by \citet{Degroote2009}, in combination with
  the comparison criterion defined in Section \ref{sec:method}, using
  $\alpha = 0.5$. We took the frequency resolution $f_{\rm res}$ as the formal error
  in the obtained frequency values. Since we included the effects of a moderate
  rotation velocity in this simulation, we searched for period spacing patterns
  by ordering the oscillation periods monotonically and by computing the
  differences between subsequent values. The results are shown in
  Fig. \ref{fig:sim_sp}, together with the spacing patterns corresponding to the
  input frequencies. Since we cut off the iterative prewhitening relatively
  quickly, we only detected a fraction of the input frequencies. This is also
  illustrated in Fig. \ref{fig:sim_comp}, where we show the fraction of detected
  input frequencies with respect to the number of prewhitened frequencies as
  well as the relative increase in the number of extracted noise frequencies. By
  using a comparison criterion with $\alpha$ = 0.5, the number of detected false
  frequencies is almost negligible. As such, the algorithm detected period
  spacings patterns at a high level of confidence.

In a second step we iteratively decreased the value of $\alpha$ from 0.4 to
0.1. This allowed us to extract more oscillation frequencies, which we could
match with the period spacings that were already detected. As shown in
Fig. \ref{fig:sim_sp}, this allowed us to extend the period spacing series without
including noise peaks.

\begin{figure}
 \includegraphics[width=88mm]{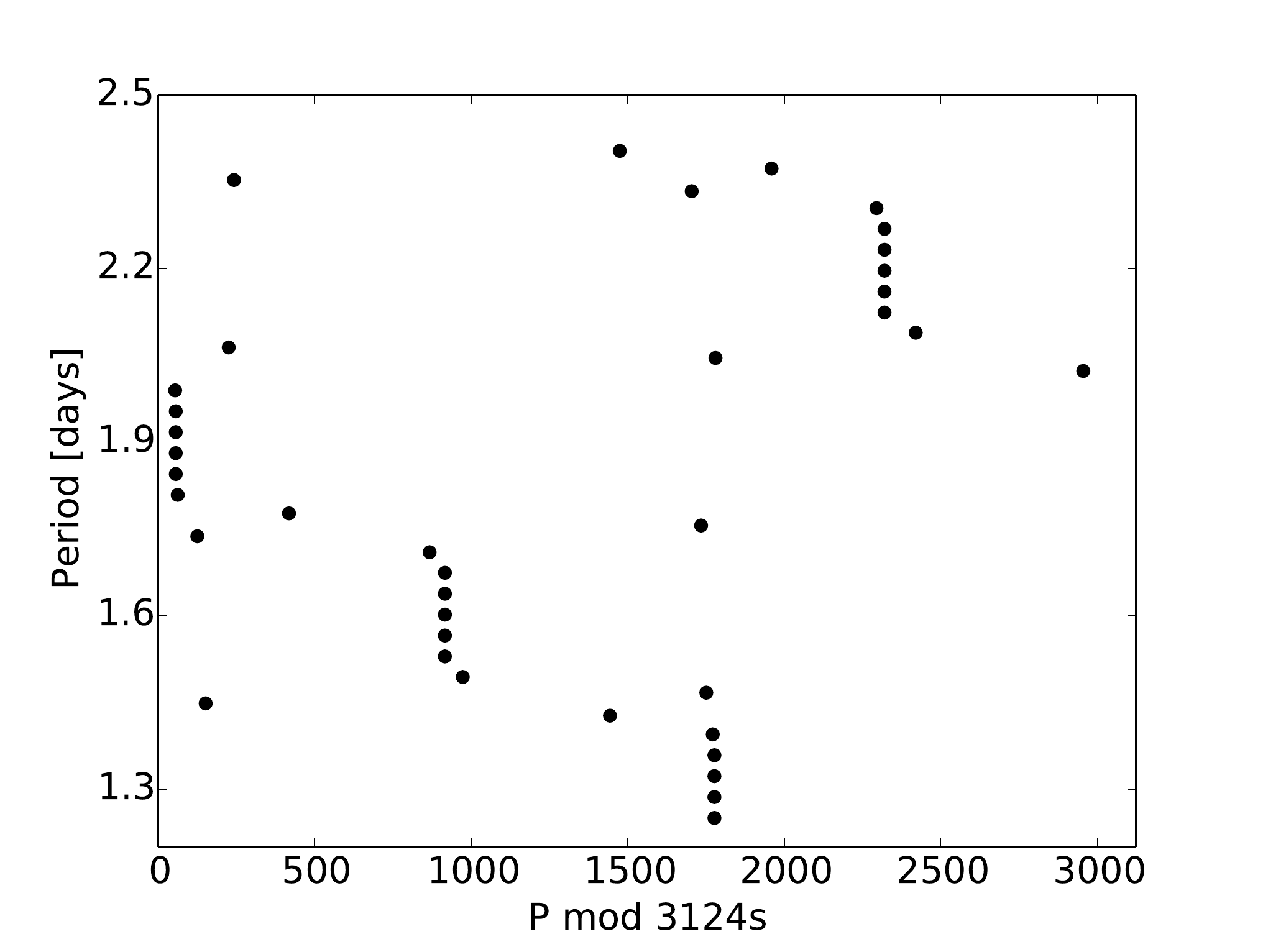}
 \caption{\label{fig:sim_echelle} Period \'echelle diagram showing the
     input frequencies of the zonal dipole modes from our simulated data.}
\end{figure}

  For slowly rotating stars, period spacing patterns
  corresponding to different $l$ and $m$ values are more likely to overlap. In
  this case, the use of \'echelle diagrams is more robust for the detection of
  period spacings. To illustrate this, we have plotted the \'echelle diagram for
  the zonal dipole modes from our previous simulation in
  Fig. \ref{fig:sim_echelle}. The zonal dipole modes have not
  undergone any rotational frequency shifts and as shown in 
  Fig. \ref{fig:sim_echelle} they form several
  ridges in the \'echelle diagram. The phase differences between these
  ridges are caused by the smaller period spacings that are 
  responsible for the dips in the period spacing pattern in
  Fig. \ref{fig:sim_sp}. The detection of such phase-shifted ridges for a star
  in an \'echelle diagram is suitable to reveal the presence of a chemical
  gradient inside the star.

\subsection{A well-studied \emph{Kepler} star: KIC\,11145123}
\label{subsec:Kurtz}
We applied our method to the hybrid $\delta$\,Sct\,/\,$\gamma$\,Dor pulsator
KIC\,11145123, which was studied by \citet{Kurtz2014}. The authors showed the
star to be a slow rotator ($P_{\rm rot}\sim 100\,\mathrm{d}$), exhibiting
almost-uniform period spacings. A large number of rotationally split multiplets
were clearly visible in the frequency spectrum, making this star a good test
case for our approach.

For this star there are about four years (Q0-Q17) of \emph{Kepler\/} data
available. Given that the light curves obtained with the standard extraction
algorithm and provided by the MAST (Mikulski Archive for Space Telescopes) were
shown to complicate the analyses for g-mode pulators and lead to a fake
instrumental low frequencies \citep{Debosscher2013,Tkachenko2013}, we applied the
  light curve extraction code developed based on customised masks by one of us
  (S.B.) to the pixel data.  The resulting light curve was then converted to
  magnitude scale and detrended for each quarter individually by subtracting a
  $1^{st}$- or $2^{nd}$-order polynomial. We concatenated the data from the
  different quarters and removed the outliers manually.

\begin{figure*}
 \includegraphics[width=\textwidth]{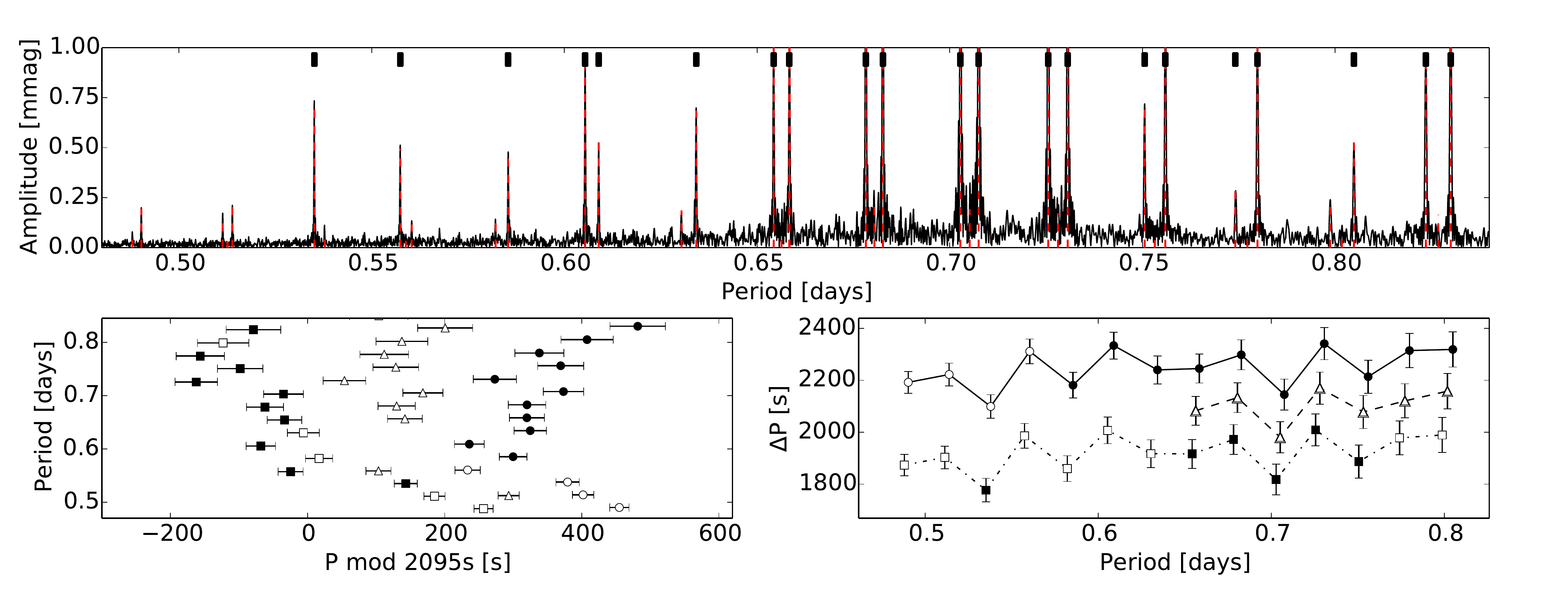}
 \caption{\label{fig:KIC11145123_sp}\textbf{Top:} A close-up of the Fourier
   transform of KIC\,11145123 (black) and the accepted oscillation
     periods (dashed red), showing the detected period spacings. The thick black
   markers indicate the frequencies that were extracted for
   $\alpha\,=\,0.5$. The amplitudes of some of the modes are $\sim$3\,mmag, but
   the Figure is zoomed in for clarity. \textbf{Bottom left:} A close-up of the
   \'echelle diagram showing the detected prograde (squares), retrograde
   (circles) and zonal (triangles) dipole modes. The frequencies that were
   extracted with $\alpha\,=\,0.5$ are marked in black. \textbf{Bottom right:}
   The detected period spacings. For clarity, we shifted the spacings of the
   prograde and retrograde modes 150\,s downwards and upwards, respectively.}
\end{figure*}

    The pulsation frequencies of KIC\,11145123 were extracted in the same
    way as for the simulated data, i.e. we performed prewhitening in
    combination with our comparison criterion described in Section
    \ref{sec:method}. Since
  this star is such a slow rotator, we used the obtained frequency
  values in combination with \'echelle diagrams to look for period spacings and
  rotational splitting. As we can see in Fig. \ref{fig:KIC11145123_sp}, we
  clearly detected part of the spacings and multiplets present in the
  data. Based on these initial results, we extracted and accepted additional
  frequencies, allowing us to extend the detected non-uniform period spacing
  series shown in Fig. \ref{fig:KIC11145123_sp}.

\begin{figure*}
 \includegraphics[width=\textwidth]{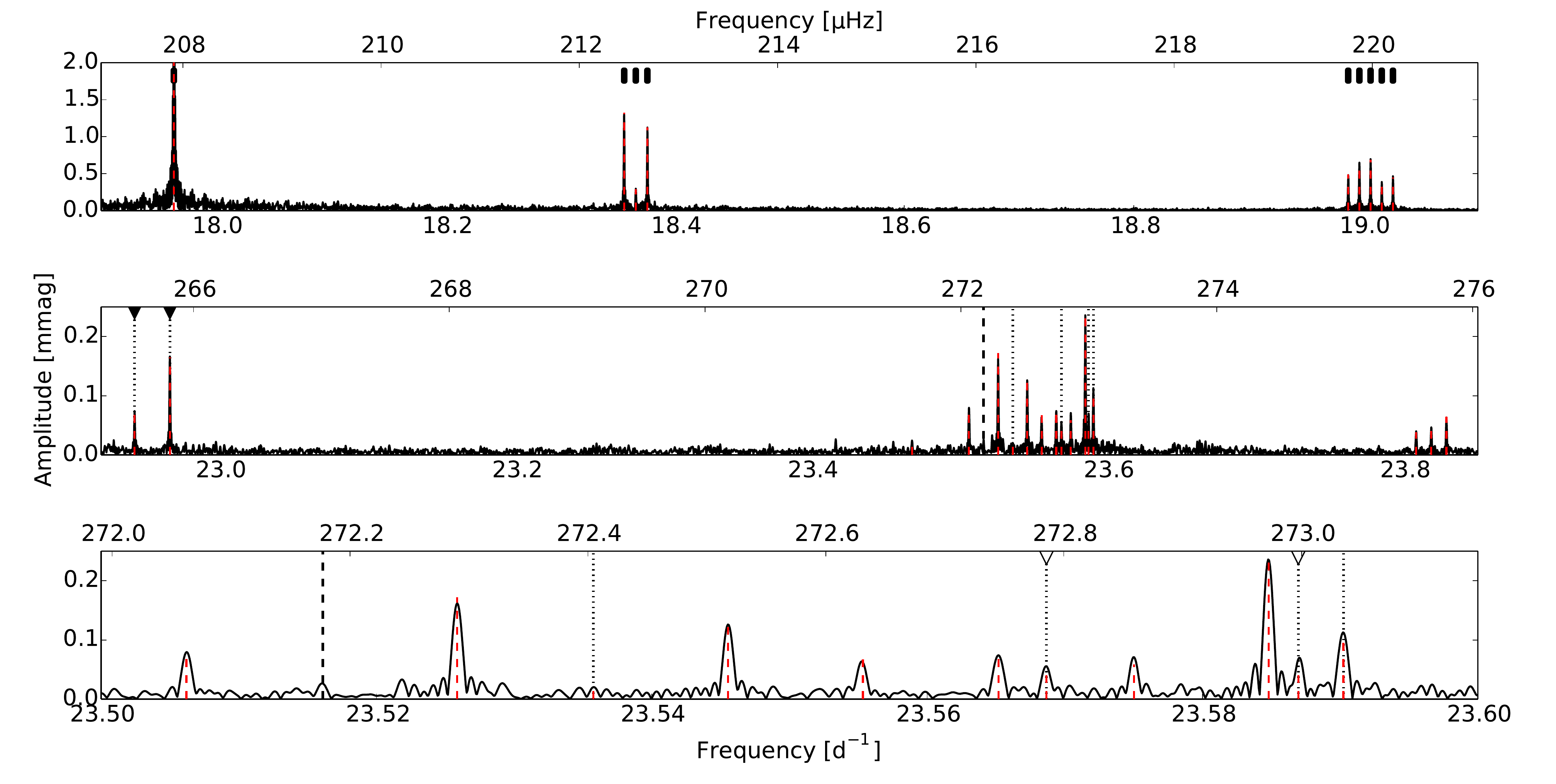}
 \caption{\label{fig:pmodes}Sections of the p-mode frequency spectrum of
   KIC\,11145123 (black), with the detected frequencies marked by a dashed red
   line. Frequencies found for $\alpha = 0.5$ are marked with thick black
   markers, whereas the one found by \citet{Kurtz2014} but not by us 
is indicated by a dashed line. The dotted lines indicate frequencies that 
we extracted, but \citet{Kurtz2014} did not. The triangles (black and
   white) indicate possible dipole doublets. The upper panel is zoomed
     in for clarity (the oscillation mode at 17.96\,d$^{-1}$ has an amplitude
     $\sim$7.2\,mmag).}
\end{figure*}

 We were able to detect the same g-mode frequencies series as found by
  \citet{Kurtz2014}, who used a classical prewhitening analysis without a stop
  criterion and considered only the 61 frequencies with the highest amplitude
  for their analysis.  We could also detect most of the p-mode multiplets
reported by these authors, as illustrated in Fig. \ref{fig:pmodes}. This is
promising, given that our method was developed with the sole aim to detect
non-uniform period spacing patterns.  We also detected two (dipole) p-mode
doublets, i.e., multiplets without a central peak, which were not listed by
\citet{Kurtz2014} (Fig.\,\ref{fig:pmodes}). On the other hand, we did
  not detect one of the p-mode triplets found by \citet{Kurtz2014} in the sense
  that one of the frequencies listed by the authors ($f$ = 23.516\,$\rm d^{-1}$)
  was not extracted by us, independently of the value for
$\alpha$. Interestingly, the non-extracted frequency is part of one of the two
multiplets which \citet{Kurtz2014} could not identify as modes with $l
\leq 2$ for their best model of 1.46\,$\rm M_{\odot}$. Instead, these were
listed as $l = 6$.  We also found an additional frequency between the
incomplete triplet and a quintuplet, as shown in the bottom plot in
Fig.\,\ref{fig:pmodes}. The inclusion of this additional frequency leads to the
detection of a larger multiplet. The extracted frequencies and splittings we
found in addition to those reported by \citet{Kurtz2014} are listed in
Table\,\ref{tab:flist} and will be used in a separate modelling study of this star.

The application of our method to the light curve of KIC\,11145123 has
  allowed us to accurately detect both the period spacing patterns and the
  rotational frequency splitting present in the data. We also found that the
presented comparison criterion might be suitable for space-based observations of
other types of variable stars, when combined with an adapted spacing pattern
detection algorithm.  However, this requires further extensive testing.

\begin{table*}
  \caption{\label{tab:flist}A least-squares fit for the additional p-mode frequencies ($*$) which we found compared to the analysis by \citet{Kurtz2014}, and for the other frequencies which are part of the same multiplets. $\Delta f$ is the spacing between the given frequency and the preceding one. The provided errors are the formal error margins from the least-squares fit.}
\begin{center}
\begin{tabular}{lccccc} \hline\hline\rule[0mm]{0mm}{3mm}
 & Frequency [$\rm d^{-1}$] & Amplitude [mmag] & Phase [2$\pi$] & $\Delta f$ [$\rm d^{-1}$] & Remarks\\
 \hline\vspace{-3mm}\\
 $*$ & 22.94259 (0.00003) & 0.074 (0.006) & -0.30 (0.08) & - & - \\ 
 $*$ & 22.96643 (0.00002) & 0.165 (0.008) &  0.46 (0.05) & 0.02383 (0.00004) & $\Delta f/2$ = 0.01192 (0.00002)\\ 
     &&&&&\\ 
     & 23.50619 (0.00003) & 0.068 (0.006) & -0.29 (0.09) & - & - \\ 
     & 23.52585 (0.00002) & 0.172 (0.008) & -0.43 (0.04) & 0.01965 (0.00004) & $\Delta f/2$ = 0.00983 (0.00002) \\ 
 $*$ & 23.53574 (0.00007) & 0.027 (0.005) &  0.47 (0.18) & 0.00990 (0.00007) & - \\ 
     & 23.54554 (0.00002) & 0.128 (0.007) &  0.37 (0.05) & 0.00980 (0.00007) & - \\ 
     & 23.55533 (0.00003) & 0.067 (0.006) &  0.46 (0.09) & 0.00979 (0.00004) & - \\ 
     & 23.56517 (0.00003) & 0.079 (0.006) & -0.07 (0.08) & 0.00984 (0.00004) & - \\ 
     & 23.57502 (0.00004) & 0.059 (0.006) &  0.18 (0.10) & 0.00984 (0.00005) & - \\ 
     & 23.58480 (0.00001) & 0.239 (0.009) & -0.16 (0.04) & 0.00978 (0.00004) & - \\ 
     &&&&&\\ 
 $*$ & 23.56866 (0.00005) & 0.045 (0.005) & 0.08 (0.12) & - & - \\ 
 $*$ & 23.58695 (0.00004) & 0.046 (0.006) & -0.39 (0.12) & 0.01829 (0.00006) & $\Delta f/2$ = 0.00915 (0.00003)\\ 
     &&&&&\\
 $*$ & 23.59023 (0.00002) & 0.109 (0.007) & -0.06 (0.06) & - & - \\
 \hline
\end{tabular}
\end{center}
\end{table*} 

\subsection{Four $\gamma$ Dor stars in the \emph{Kepler} field}
\label{subsec:Kepler}
We also applied the proposed method to four of 
the $\gamma$ Doradus pulsators from the
sample presented by \citet{Tkachenko2013}, for which we have both
high-resolution spectroscopy and \emph{Kepler} photometry at our disposal.
For each of those four stars we have obtained four high-resolution spectra with the
HERMES spectrograph \citep[377-900\,nm, R$\sim$85000,][]{Raskin2011} at the
1.2-m Mercator telescope (Observatorio del Roque de los Muchachos, La Palma,
Canary Islands). These spectra were reduced with the dedicated HERMES pipeline,
and subsequently normalised following the method described by
\citet{Papics2012}.  We used the improved LSD algorithm \citep{Tkachenko2013b}
to compute high S/N average profiles from these spectroscopic observations. The
LSD profiles were then checked for the presence of a binary companion and line
profile variations, which confirmed that the four stars are single objects
(see Table \ref{tab:param}).  We computed an average spectrum and analysed it to
determine the fundamental and atmospheric parameters by means of the spectrum
synthesis method implemented in the GSSP code
\citep{Lehmann2013,Tkachenko2013c}. The advantages of using the average spectra
are that (i) they have higher signal-to-noise ratios than the individual
spectra, and (ii) line profile deformations due to the pulsations are largely
smoothed out. The obtained values are listed in Table \ref{tab:param} and place
the four stars in the $\gamma$ Doradus strip.

\begin{table*}
  \caption{\label{tab:param}Atmospheric parameter values obtained for the four $\gamma$ Dor stars analysed in this study. The provided error bars are the 1-$\sigma$ values obtained from the chi-square statistics.}
\begin{center}
\begin{tabular}{lccccc} \hline\hline\rule[0mm]{0mm}{3mm}
 Kepler ID & $T_{\rm eff}$ & $\log\,g$ & $[M/H]$ & $v\sin\,i$ & $\xi$\\
  & [$K$] & [dex] & [dex] & [$\rm km\,s^{-1}$] & [$\rm km\,s^{-1}$]\\
 \hline\vspace{-3mm}\\
 KIC\,5350598 & 7200 (100) & 3.83 (0.35) & -0.21 (0.10) & 26.2 (1.7) & 3.0 (0.5)\\
 KIC\,6185513 & 7300 (200) & 4.50 (0.75) & -0.07 (0.17) & 78.0 (11.0) & 3.0 (1.7)\\
 KIC\,6678174 & 7200 (100) & 4.00 (0.42) & -0.19 (0.11) & 43.6 (2.8) & 2.9 (0.6)\\
 KIC\,11721304 & 7300 (100) & 4.42 (0.32) & 0.01 (0.10) & 27.4 (1.6) & 2.3 (0.4)\\
 \hline
\end{tabular}
\end{center}
\end{table*} 

For KIC\,5350598 and KIC\,11721304 about four years of photometric 
observations (Q0-Q17) were obtained with {\it Kepler}, while for KIC\,6185513 
(Q0-Q1, Q10-Q12, Q14-Q16) and KIC\,6778174 (Q0-Q1, Q10-Q17) about two years of observations  
are available. The pixel data were extracted, reduced and analysed
in the same way as the data for KIC\,11145123, as described in Section
\ref{subsec:Kurtz}, and unless mentioned otherwise, $\alpha$ was taken to be
0.5. Using the obtained frequency values to compute the period spacing, we
discuss the patterns for each star individually.

\subsubsection{KIC\,11721304}
\label{subsubsec:KIC11721304}
\begin{figure*}
 \includegraphics[width=\textwidth]{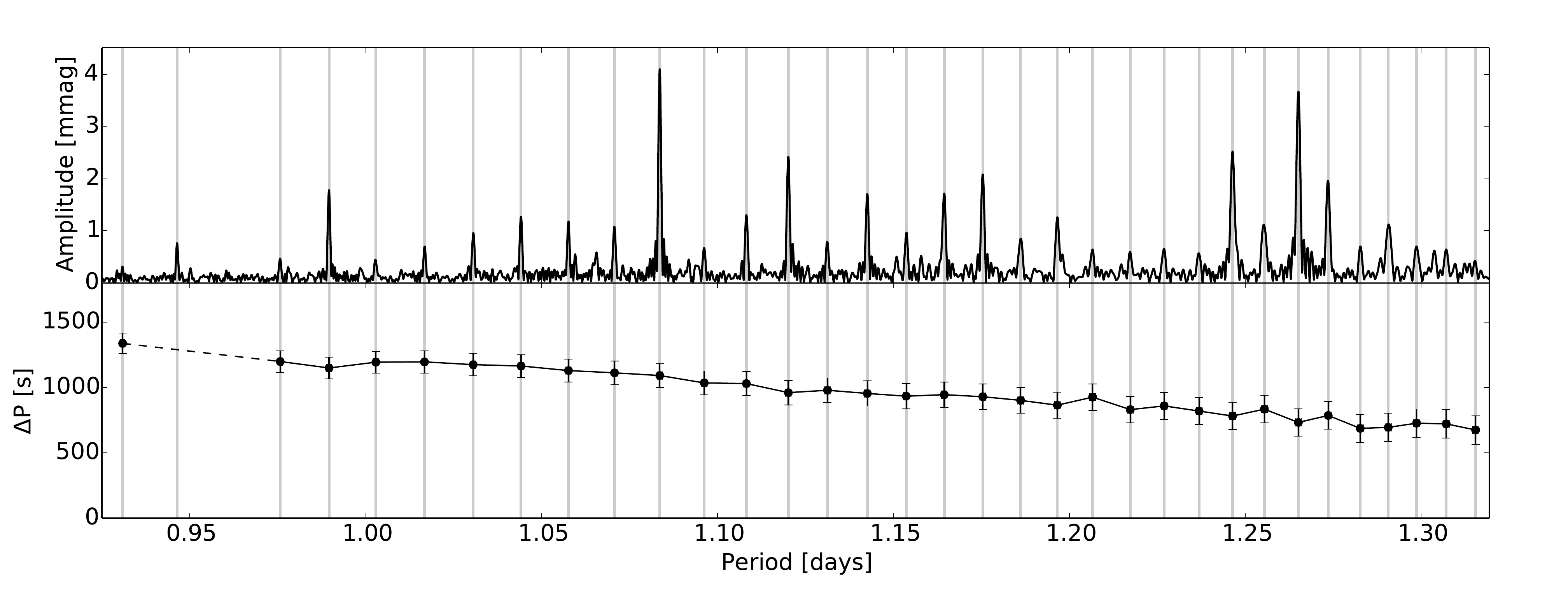}
 \caption{\label{fig:KIC11721304_perind}\textbf{Top:} A close-up of part of the
   Fourier spectrum of KIC\,11721304 (black). All the marked frequencies
     are accepted following the criterion described in Section \ref{sec:method},
     with $\alpha$ = 0.5. \textbf{Bottom:} The period spacing as computed from
     the accepted frequencies. The black markers and grey lines indicate the
     frequencies for which we find a smooth spacing pattern.}
\end{figure*}

For KIC\,11721304 we have detected a 
period spacing pattern with a clear slope, 
as shown in Fig. \ref{fig:KIC11721304_perind}. 
The average period spacing (1200\,s) is
  small compared to typical values for a non-rotating
  $\gamma\,$Doradus star. These are of the order of 3000\,s ($l$ = 1) to
  2000\,s ($l$ = 2), which implies that the pulsations in the detected series are
  prograde modes. The downward trend of the spacing pattern supports this
  interpretation.
 
  In addition to the pulsation frequencies belonging to this spacing pattern,
  we also detected additional frequencies in the same range.  These likely
  correspond with modes of different wavenumbers $l$ and $m$, but firm
  conclusions about this requires detailed future modelling, taking into account
  various mixing processes. However, this kind of modelling is outside of the 
  cope of the current research.
 
\begin{figure}
 \includegraphics[width=88mm]{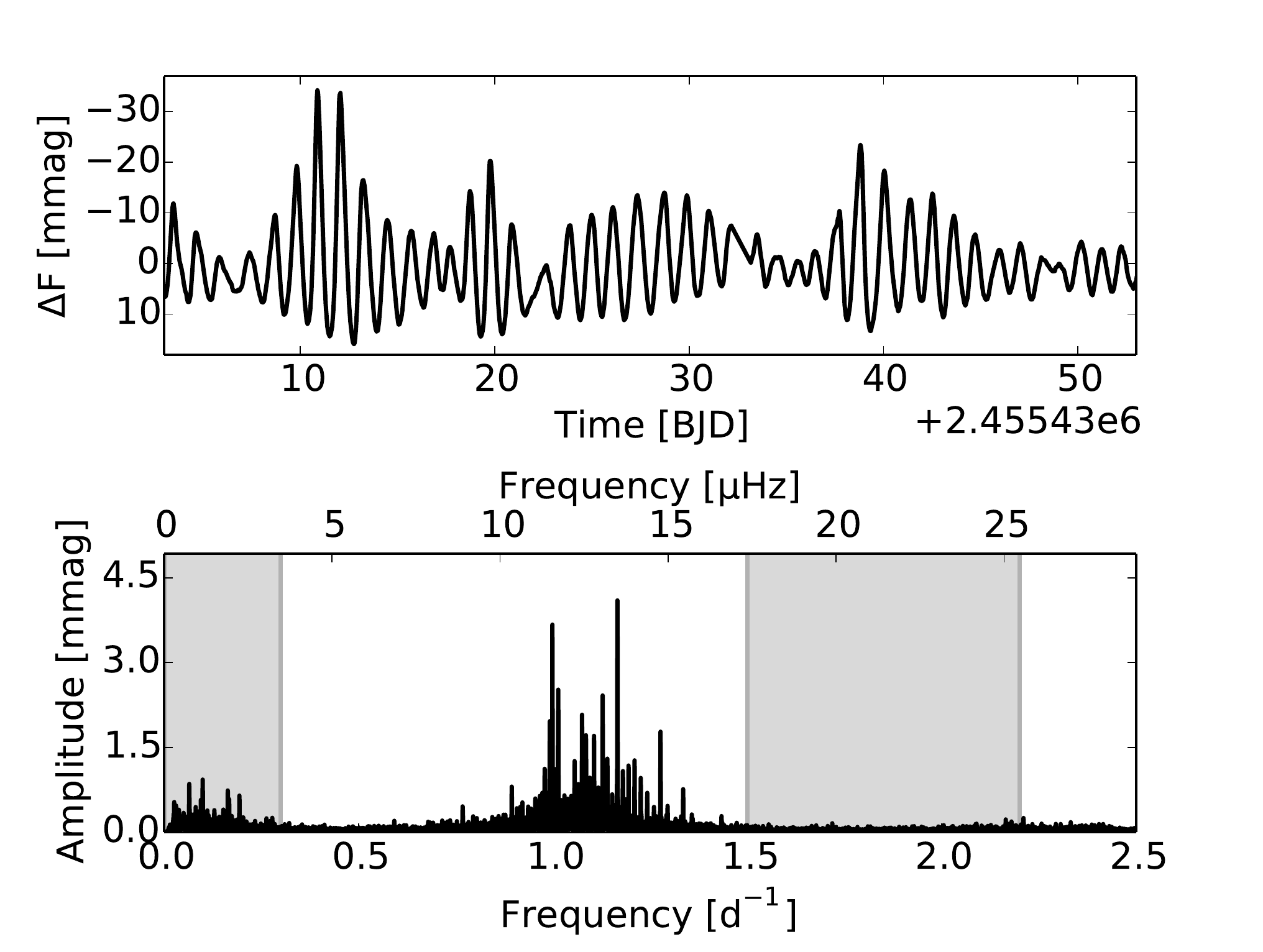}
 \caption{\label{fig:KIC11721304_sim} \textbf{Top:} A small part of the
     light curve of KIC\,11721304. \textbf{Bottom:} The Fourier spectrum (black)
     of KIC\,11721304. The grey areas indicate the locations of the linear
     combinations of the oscillation frequencies, which arise because of the
     asymmetry of the light curve of KIC\,11721304.}
\end{figure}

  We saw in Section\,\ref{sec:method} that many $\gamma$ Doradus
  stars have asymmetric light curves and that this leads to the appearance of
  linear combinations of the frequencies in their Fourier spectra. In general, we
  have to make sure that we did not include any of these combination frequencies
  in the period spacing series. This is not an issue in the case of
  KIC\,11721304 because the combination frequencies are located sufficiently
  far from the detected period spacing series in the frequency domain 
  (Fig. \ref{fig:KIC11721304_sim}).

\begin{figure}
 \includegraphics[width=88mm]{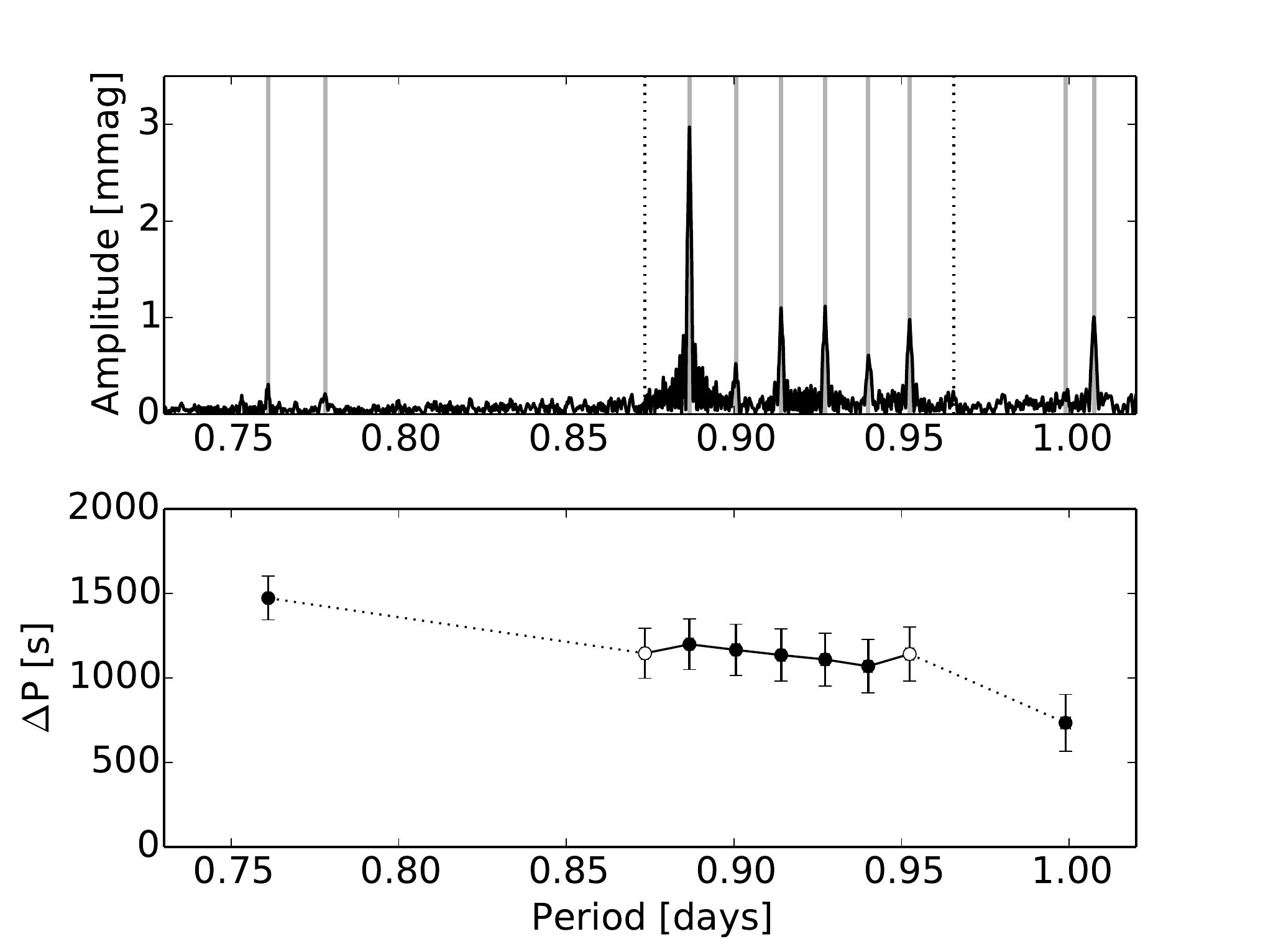}
 \caption{\label{fig:KIC6678174_sp}\textbf{Top:} A close-up of the Fourier
   transform of the light curve of KIC\,6678174 (black) and the accepted
   frequencies in the region where the detected period spacing series is
   located. The frequencies which were accepted for $\alpha\,=\,0.5$ are marked
   with full grey lines, and the ones for which $\alpha\,<\,0.5$ are marked with
   dotted lines. \textbf{Bottom:} The period spacing as computed for the
   (accepted) frequencies shown in the top. The white markers indicate the
   spacings for frequencies accepted for $\alpha\,<\,0.5$, while the dotted
   lines indicate the regions where there are likely frequencies missing.}
\end{figure}

\subsubsection{KIC\,6678174}
\label{subsubsec:KIC6678174}
In the case of KIC\,6678174, there is also a clear detection of a period spacing
series, though the length of the found series is relatively short
(Fig.\,\ref{fig:KIC6678174_sp}). Similar to the period spacing series we
  detected for KIC\,11721304, there are no dips present in the pattern and the
  value of the average spacing is similar.  This seems to point to a common
  property in our sample of $\gamma\,$Doradus stars \citep{Tkachenko2013},
  although it requires the analysis of the full sample to make any firm
  conclusions. Such extensive analysis study will be undertaken in the near
  future as a follow-up of the current methodological paper.

\subsubsection{KIC\,5350598 and KIC\,6185513}
\label{subsubsec:nodetection}
For KIC\,5350598 and KIC\,6185513, we could not detect a period spacing
pattern. The analyses suggest that the number of resolved oscillation
frequencies is insufficient for these two stars to allow such a pattern.  For
KIC\,5350598 we found frequencies that might be part of a spacing pattern, but
there are too many missing frequencies in the series to arrive at a firm
conclusion. These missing frequencies could be due excitation with lower
amplitude than the other modes that could constitute a pattern but the present
data set does not allow a firm conclusion.

For KIC\,6185513, we arrive at a candidate spacing pattern with $\Delta
P\,\sim400\,\mathrm{s}$ for $\alpha\,\leq\,0.3$. Even though this is consistent
with the star's $v\sin i$ value ($78\pm11\,\mathrm{km\,s^{-1}}$), we treat the
result with caution because the pattern consists of only eight frequencies and
they are not consecutive as for the case of KIC\,11721304.

\section{Discussion and conclusions}
\label{sec:conclusions}

 In this study we have developed a method to detect reliable period
  spacings in the complex frequency spectra of $\gamma$ Doradus stars. The
  various steps are as follows:
 \begin{enumerate}
  \item Extract the frequencies from the light curve using a prewhitening
    method.
  \item Adopt a conservative stop criterion for the prewhitening procedure; we
    propose to use what we termed the comparison criterion given in
    Eq.\,(\ref{stop}), where $\alpha$ = 0.5 was found to be a suitable value for
    $\gamma$ Doradus stars for the first iteration.
  \item Make a distinction between moderate to fast versus slow rotators when
    searching for the period spacing patterns. For the former, order the
    extracted oscillation periods monotonically and compute the differences
    between subsequent values to look for period spacings. For slow rotators,
    \'echelle diagrams were found to be the most robust method to search for
    period spacing patterns.
  \item Iterate over the method while lowering the value for $\alpha$, allowing
    us to extract more frequencies from the data. If period spacings were already
    detected, use these as a guide to check which additional frequency values
    are reliable and to avoid the influence of frequencies due to noise. If no
    period spacings were detected yet, the extraction of additional frequency
    peaks at lower $\alpha$ could lead to a detection, though it is important to
    keep in mind that the detection of false frequencies due to noise also
    becomes more likely.
\end{enumerate}

In the present study, we have taken a closer look at the results for
KIC\,11145123, a hybrid pulsator recently presented by \citet{Kurtz2014}, and at
four single $\gamma\,$Doradus stars from the sample presented by
\citet{Tkachenko2013}. From the application of our method on KIC\,11145123 we
deduced that the method excels in the exclusion of noise peaks. Because of this
star's slow rotation, we detected the frequency splittings and the period
spacings using \'echelle diagrams.  For two of the other four stars we
studied, KIC\,6678174 and KIC\,11721304, we were able to detect period spacings
that are completely in line with theoretical expectations as presented in
\citet{Miglio2008a} and \citet{Bouabid2013}. The analysis of KIC\,11721304 was
particularly successful, resulting in a series of 35 pulsation periods, hinting
towards strong mixing processes. The frequency analysis of the two
  additional stars did not lead to firm conclusions.

  It is important to keep in mind that the method cannot be automated and
  it remains necessary for the user to check carefully any detected period
  spacing pattern. As such, the applicant should be familiar with the basic
  aspects of the theoretical expectations for non-uniform period spacings.
  Detected spacings should then be verified and interpreted in terms of interior
  physics using extensive grids of theoretical pulsation properties from a
  combination of stellar evolution and pulsation codes. Such modelling will also
  allow us to assign values to the pulsation parameters $l$ and $m$.

\begin{acknowledgements}
  The research leading to these results was based on funding from the Fund for
  Scientific Research of Flanders (FWO), Belgium, under grant agreement
  G.0B69.13 and from the European Community's Seventh Framework Programme
  FP7-SPACE-2011-1, project number 312844 (SPACEINN).  PP has received funding 
  from the Research Fund KU Leuven. SB is supported by the Foundation for 
  Fundamental Research on Matter (FOM), which is part of the Netherlands 
  Organisation for Scientific Research (NWO). Funding for the \emph{Kepler} 
  mission is provided by NASA's Science Mission Directorate. We thank the whole 
  team for the development and operations of the mission. This research made use 
  of the SIMBAD database and the VizieR catalogue access tool operated at CDS, 
  Strasbourg, France, and the SAO/NASA Astrophysics Data System.
\end{acknowledgements}

\bibliographystyle{aa}
\bibliography{gDor_methodology}

\begin{thebibliography}{48}
\expandafter\ifx\csname natexlab\endcsname\relax\def\natexlab#1{#1}\fi

\bibitem[{{Aerts} {et~al.}(2010){Aerts}, {Christensen-Dalsgaard}, \&
  {Kurtz}}]{Aerts2010}
{Aerts}, C., {Christensen-Dalsgaard}, J., \& {Kurtz}, D.~W. 2010,
  {Asteroseismology, Astronomy and Astrophsyics Library, Springer Berlin
  Heidelberg}

\bibitem[{{Auvergne} {et~al.}(2009){Auvergne}, {Bodin}, {Boisnard}, {Buey},
  {Chaintreuil}, {Epstein}, {Jouret}, {Lam-Trong}, {Levacher}, {Magnan},
  {Perez}, {Plasson}, {Plesseria}, {Peter}, {Steller}, {Tiph{\`e}ne}, {Baglin},
  {Agogu{\'e}}, {Appourchaux}, {Barbet}, {Beaufort}, {Bellenger}, {Berlin},
  {Bernardi}, {Blouin}, {Boumier}, {Bonneau}, {Briet}, {Butler}, {Cautain},
  {Chiavassa}, {Costes}, {Cuvilho}, {Cunha-Parro}, {de Oliveira Fialho},
  {Decaudin}, {Defise}, {Djalal}, {Docclo}, {Drummond}, {Dupuis}, {Exil},
  {Faur{\'e}}, {Gaboriaud}, {Gamet}, {Gavalda}, {Grolleau}, {Gueguen},
  {Guivarc'h}, {Guterman}, {Hasiba}, {Huntzinger}, {Hustaix}, {Imbert},
  {Jeanville}, {Johlander}, {Jorda}, {Journoud}, {Karioty}, {Kerjean},
  {Lafond}, {Lapeyrere}, {Landiech}, {Larqu{\'e}}, {Laudet}, {Le Merrer},
  {Leporati}, {Leruyet}, {Levieuge}, {Llebaria}, {Martin}, {Mazy}, {Mesnager},
  {Michel}, {Moalic}, {Monjoin}, {Naudet}, {Neukirchner}, {Nguyen-Kim},
  {Ollivier}, {Orcesi}, {Ottacher}, {Oulali}, {Parisot}, {Perruchot},
  {Piacentino}, {Pinheiro da Silva}, {Platzer}, {Pontet}, {Pradines},
  {Quentin}, {Rohbeck}, {Rolland}, {Rollenhagen}, {Romagnan}, {Russ}, {Samadi},
  {Schmidt}, {Schwartz}, {Sebbag}, {Smit}, {Sunter}, {Tello}, {Toulouse},
  {Ulmer}, {Vandermarcq}, {Vergnault}, {Wallner}, {Waultier}, \&
  {Zanatta}}]{Auvergne2009}
{Auvergne}, M., {Bodin}, P., {Boisnard}, L., {et~al.} 2009, A\&A, 506, 411

\bibitem[{{Balona}(2012)}]{Balona2012}
{Balona}, L.~A. 2012, \mnras, 422, 1092

\bibitem[{{Balona}(2014)}]{Balona2014}
{Balona}, L.~A. 2014, \mnras, 439, 3453

\bibitem[{{Balona} {et~al.}(2011){Balona}, {Guzik}, {Uytterhoeven}, {Smith},
  {Tenenbaum}, \& {Twicken}}]{Balona2011}
{Balona}, L.~A., {Guzik}, J.~A., {Uytterhoeven}, K., {et~al.} 2011, \mnras,
  415, 3531

\bibitem[{{Beck} {et~al.}(2011){Beck}, {Bedding}, {Mosser}, {Stello}, {Garcia},
  {Kallinger}, {Hekker}, {Elsworth}, {Frandsen}, {Carrier}, {De Ridder},
  {Aerts}, {White}, {Huber}, {Dupret}, {Montalb{\'a}n}, {Miglio}, {Noels},
  {Chaplin}, {Kjeldsen}, {Christensen-Dalsgaard}, {Gilliland}, {Brown},
  {Kawaler}, {Mathur}, \& {Jenkins}}]{Beck2011}
{Beck}, P.~G., {Bedding}, T.~R., {Mosser}, B., {et~al.} 2011, Science, 332, 205

\bibitem[{{Beck} {et~al.}(2012){Beck}, {Montalban}, {Kallinger}, {De Ridder},
  {Aerts}, {Garc{\'i}a}, {Hekker}, {Dupret}, {Mosser}, {Eggenberger}, {Stello},
  {Elsworth}, {Frandsen}, {Carrier}, {Hillen}, {Gruberbauer},
  {Christensen-Dalsgaard}, {Miglio}, {Valentini}, {Bedding}, {Kjeldsen},
  {Girouard}, {Hall}, \& {Ibrahim}}]{Beck2012}
{Beck}, P.~G., {Montalban}, J., {Kallinger}, T., {et~al.} 2012, Nat., 481, 55

\bibitem[{{Bedding} {et~al.}(2011){Bedding}, {Mosser}, {Huber},
  {Montalb{\'a}n}, {Beck}, {Christensen-Dalsgaard}, {Elsworth}, {Garc{\'i}a},
  {Miglio}, {Stello}, {White}, {De Ridder}, {Hekker}, {Aerts}, {Barban},
  {Belkacem}, {Broomhall}, {Brown}, {Buzasi}, {Carrier}, {Chaplin}, {di Mauro},
  {Dupret}, {Frandsen}, {Gilliland}, {Goupil}, {Jenkins}, {Kallinger},
  {Kawaler}, {Kjeldsen}, {Mathur}, {Noels}, {Aguirre}, \&
  {Ventura}}]{Bedding2011}
{Bedding}, T.~R., {Mosser}, B., {Huber}, D., {et~al.} 2011, \nat, 471, 608

\bibitem[{{Bouabid} {et~al.}(2013){Bouabid}, {Dupret}, {Salmon},
  {Montalb{\'a}n}, {Miglio}, \& {Noels}}]{Bouabid2013}
{Bouabid}, M.-P., {Dupret}, M.-A., {Salmon}, S., {et~al.} 2013, \mnras, 429,
  2500

\bibitem[{{Breger} {et~al.}(2012){Breger}, {Fossati}, {Balona}, {Kurtz},
  {Robertson}, {Bohlender}, {Lenz}, {M{\"u}ller}, {L{\"u}ftinger}, {Clarke},
  {Hall}, \& {Ibrahim}}]{Breger2012}
{Breger}, M., {Fossati}, L., {Balona}, L., {et~al.} 2012, \apj, 759, 62

\bibitem[{{Breger} {et~al.}(1993){Breger}, {Stich}, {Garrido}, {Martin},
  {Jiang}, {Li}, {Hube}, {Ostermann}, {Paparo}, \& {Scheck}}]{Breger1993}
{Breger}, M., {Stich}, J., {Garrido}, R., {et~al.} 1993, \aap, 271, 482

\bibitem[{{Brunsden} {et~al.}(2012){Brunsden}, {Pollard}, {Cottrell}, {Wright},
  \& {De Cat}}]{Brunsden2012}
{Brunsden}, E., {Pollard}, K.~R., {Cottrell}, P.~L., {Wright}, D.~J., \& {De
  Cat}, P. 2012, Mon.Not.R.Astron.Soc., 427, 2512

\bibitem[{{Chadid} {et~al.}(2001){Chadid}, {De Ridder}, {Aerts}, \&
  {Mathias}}]{Chadid2001}
{Chadid}, M., {De Ridder}, J., {Aerts}, C., \& {Mathias}, P. 2001, \aap, 375,
  113

\bibitem[{{Chapellier} {et~al.}(2012){Chapellier}, {Mathias}, {Weiss}, {Le
  Contel}, \& {Debosscher}}]{Chapellier2012}
{Chapellier}, E., {Mathias}, P., {Weiss}, W.~W., {Le Contel}, D., \&
  {Debosscher}, J. 2012, \aap, 540, A117

\bibitem[{{Chlebowski}(1978)}]{Chlebowski1978}
{Chlebowski}, T. 1978, \actaa, 28, 441

\bibitem[{{Cuypers} {et~al.}(2009){Cuypers}, {Aerts}, {De Cat}, {De Ridder},
  {Goossens}, {Schoenaers}, {Uytterhoeven}, {Acke}, {Davignon}, {Debosscher},
  {Decin}, {De Meester}, {Deroo}, {Drummond}, {Kolenberg}, {Lefever}, {Raskin},
  {Reyniers}, {Saesen}, {Vandenbussche}, {van Malderen}, {Verhoelst}, {van
  Winckel}, \& {Waelkens}}]{Cuypers2009}
{Cuypers}, J., {Aerts}, C., {De Cat}, P., {et~al.} 2009, \aap, 499, 967

\bibitem[{{Davie} {et~al.}(2014){Davie}, {Pollard}, {Cottrell}, {Brunsden},
  {Wright}, \& {De Cat}}]{Davie2014}
{Davie}, M.~W., {Pollard}, K.~R., {Cottrell}, P.~L., {et~al.} 2014, \pasa, 31,
  25

\bibitem[{{Debosscher} {et~al.}(2013){Debosscher}, {Aerts}, {Tkachenko},
  {Pavlovski}, {Maceroni}, {Kurtz}, {Beck}, {Bloemen}, {Degroote}, {Lombaert},
  \& {Southworth}}]{Debosscher2013}
{Debosscher}, J., {Aerts}, C., {Tkachenko}, A., {et~al.} 2013, \aap, 556, A56

\bibitem[{{Degroote} {et~al.}(2010){Degroote}, {Aerts}, {Baglin}, {Miglio},
  {Briquet}, {Noels}, {Niemczura}, {Montalban}, {Bloemen}, {Oreiro},
  {Vu\v{c}kovi{\'c}}, {Smolders}, {Auvergne}, {Baudin}, {Catala}, \&
  {Michel}}]{Degroote2010}
{Degroote}, P., {Aerts}, C., {Baglin}, A., {et~al.} 2010, \nat, 464, 259

\bibitem[{{Degroote} {et~al.}(2009){Degroote}, {Briquet}, {Catala},
  {Uytterhoeven}, {Lefever}, {Morel}, {Aerts}, {Carrier}, {Auvergne}, {Baglin},
  \& {Michel}}]{Degroote2009}
{Degroote}, P., {Briquet}, M., {Catala}, C., {et~al.} 2009, \aap, 506, 111

\bibitem[{{Deheuvels} {et~al.}(2014){Deheuvels}, {Do\u{g}an}, {Goupil},
  {Appourchaux}, {Benomar}, {Bruntt}, {Campante}, {Casagrande}, {Ceillier},
  {Davies}, {De Cat}, {Fu}, {Garc{\'i}a}, {Lobel}, {Mosser}, {Reese}, {Regulo},
  {Schou}, {Stahn}, {Thygesen}, {Yang}, {Chaplin}, {Christensen-Dalsgaard},
  {Eggenberger}, {Gizon}, {Mathis}, {Molenda-{\.Z}akowicz}, \&
  {Pinsonneault}}]{Deheuvels2014}
{Deheuvels}, S., {Do\u{g}an}, G., {Goupil}, M.~J., {et~al.} 2014, \aap, 564,
  A27

\bibitem[{{Dupret} {et~al.}(2005){Dupret}, {Grigahc{\`e}ne}, {Garrido},
  {Gabriel}, \& {Scuflaire}}]{Dupret2005}
{Dupret}, M.-A., {Grigahc{\`e}ne}, A., {Garrido}, R., {Gabriel}, M., \&
  {Scuflaire}, R. 2005, \aap, 435, 927

\bibitem[{{Grec} {et~al.}(1983){Grec}, {Fossat}, \& {Pomerantz}}]{Grec1983}
{Grec}, G., {Fossat}, E., \& {Pomerantz}, M.~A. 1983, \solphys, 82, 55

\bibitem[{{Guzik} {et~al.}(2000){Guzik}, {Kaye}, {Bradley}, {Cox}, \&
  {Neuforge}}]{Guzik2000}
{Guzik}, J.~A., {Kaye}, A.~B., {Bradley}, P.~A., {Cox}, A.~N., \& {Neuforge},
  C. 2000, \apjl, 542, L57

\bibitem[{{Handler}(1999)}]{Handler1999}
{Handler}, G. 1999, \mnras, 309, L19

\bibitem[{{Hareter}(2012)}]{Hareter2012}
{Hareter}, M. 2012, Astronomische Nachrichten, 333, 1048

\bibitem[{{Herwig}(2000)}]{Herwig2000}
{Herwig}, F. 2000, \aap, 360, 952

\bibitem[{{Kaye} {et~al.}(1999){Kaye}, {Handler}, {Krisciunas}, {Poretti}, \&
  {Zerbi}}]{Kaye1999}
{Kaye}, A.~B., {Handler}, G., {Krisciunas}, K., {Poretti}, E., \& {Zerbi},
  F.~M. 1999, \pasp, 111, 840

\bibitem[{{Koch} {et~al.}(2010){Koch}, {Borucki}, {Basri}, {Batalha}, {Brown},
  {Caldwell}, {Christensen-Dalsgaard}, {Cochran}, {DeVore}, {Dunham},
  {Gautier}, {Geary}, {Gilliland}, {Gould}, {Jenkins}, {Kondo}, {Latham},
  {Lissauer}, {Marcy}, {Monet}, {Sasselov}, {Boss}, {Brownlee}, {Caldwell},
  {Dupree}, {Howell}, {Kjeldsen}, {Meibom}, {Morrison}, {Owen}, {Reitsema},
  {Tarter}, {Bryson}, {Dotson}, {Gazis}, {Haas}, {Kolodziejczak}, {Rowe}, {Van
  Cleve}, {Allen}, {Chandrasekaran}, {Clarke}, {Li}, {Quintana}, {Tenenbaum},
  {Twicken}, \& {Wu}}]{Koch2010}
{Koch}, D.~G., {Borucki}, W.~J., {Basri}, G., {et~al.} 2010, ApJ, 713, L79

\bibitem[{{Kurtz} {et~al.}(2014){Kurtz}, {Saio}, {Takata}, {Shibahashi},
  {Murphy}, \& {Sekii}}]{Kurtz2014}
{Kurtz}, D.~W., {Saio}, H., {Takata}, M., {et~al.} 2014, \mnras, 444, 102

\bibitem[{{Lehmann} {et~al.}(2013){Lehmann}, {Southworth}, {Tkachenko}, \&
  {Pavlovski}}]{Lehmann2013}
{Lehmann}, H., {Southworth}, J., {Tkachenko}, A., \& {Pavlovski}, K. 2013,
  \aap, 557, A79

\bibitem[{{Miglio} {et~al.}(2008){Miglio}, {Montalb{\'a}n}, {Noels}, \&
  {Eggenberger}}]{Miglio2008a}
{Miglio}, A., {Montalb{\'a}n}, J., {Noels}, A., \& {Eggenberger}, P. 2008,
  \mnras, 386, 1487

\bibitem[{{Mosser} {et~al.}(2012){Mosser}, {Goupil}, {Belkacem}, {Marques},
  {Beck}, {Bloemen}, {De Ridder}, {Barban}, {Deheuvels}, {Elsworth}, {Hekker},
  {Kallinger}, {Ouazzani}, {Pinsonneault}, {Samadi}, {Stello}, {Garc{\'i}a},
  {Klaus}, {Li}, {Mathur}, \& {Morris}}]{Mosser2012}
{Mosser}, B., {Goupil}, M.~J., {Belkacem}, K., {et~al.} 2012, \aap, 548, A10

\bibitem[{{Mosser} {et~al.}(2013){Mosser}, {Michel}, {Belkacem}, {Goupil},
  {Baglin}, {Barban}, {Provost}, {Samadi}, {Auvergne}, \&
  {Catala}}]{Mosser2013}
{Mosser}, B., {Michel}, E., {Belkacem}, K., {et~al.} 2013, \aap, 550, A126

\bibitem[{{P{\'a}pics}(2012)}]{Papics2012b}
{P{\'a}pics}, P.~I. 2012, Astronomische Nachrichten, 333, 1053

\bibitem[{{P{\'a}pics} {et~al.}(2012){P{\'a}pics}, {Briquet}, {Baglin},
  {Poretti}, {Aerts}, {Degroote}, {Tkachenko}, {Morel}, {Zima}, {Niemczura},
  {Rainer}, {Hareter}, {Baudin}, {Catala}, {Michel}, {Samadi}, \&
  {Auvergne}}]{Papics2012}
{P{\'a}pics}, P.~I., {Briquet}, M., {Baglin}, A., {et~al.} 2012, \aap, 542, A55

\bibitem[{{P{\'a}pics} {et~al.}(2014){P{\'a}pics}, {Moravveji}, {Aerts},
  {Tkachenko}, {Triana}, {Bloemen}, \& {Southworth}}]{Papics2014}
{P{\'a}pics}, P.~I., {Moravveji}, E., {Aerts}, C., {et~al.} 2014, \aap, in
  press

\bibitem[{{Paxton} {et~al.}(2011){Paxton}, {Bildsten}, {Dotter}, {Herwig},
  {Lesaffre}, \& {Timmes}}]{Paxton2011}
{Paxton}, B., {Bildsten}, L., {Dotter}, A., {et~al.} 2011, \apjs, 192, 3

\bibitem[{{Paxton} {et~al.}(2013){Paxton}, {Cantiello}, {Arras}, {Bildsten},
  {Brown}, {Dotter}, {Mankovich}, {Montgomery}, {Stello}, {Timmes}, \&
  {Townsend}}]{Paxton2013}
{Paxton}, B., {Cantiello}, M., {Arras}, P., {et~al.} 2013, \apjs, 208, 4

\bibitem[{{Raskin} {et~al.}(2011){Raskin}, {van Winckel}, {Hensberge},
  {Jorissen}, {Lehmann}, {Waelkens}, {Avila}, {de Cuyper}, {Degroote},
  {Dubosson}, {Dumortier}, {Fr{\'e}mat}, {Laux}, {Michaud}, {Morren}, {Perez
  Padilla}, {Pessemier}, {Prins}, {Smolders}, {van Eck}, \&
  {Winkler}}]{Raskin2011}
{Raskin}, G., {van Winckel}, H., {Hensberge}, H., {et~al.} 2011, A\&A, 526, A69

\bibitem[{{Reed} {et~al.}(2011){Reed}, {Baran}, {Quint}, {Kawaler}, {O'Toole},
  {Telting}, {Charpinet}, {Rodr{\'i}guez-L{\'o}pez}, {{\O}stensen},
  {Provencal}, {Johnson}, {Thompson}, {Allen}, {Middour}, {Kjeldsen}, \&
  {Christensen-Dalsgaard}}]{Reed2011}
{Reed}, M.~D., {Baran}, A., {Quint}, A.~C., {et~al.} 2011, \mnras, 414, 2885

\bibitem[{{Savonije}(2013)}]{Savonije2013}
{Savonije}, G.~J. 2013, \aap, 559, A25

\bibitem[{{Tassoul}(1980)}]{Tassoul1980}
{Tassoul}, M. 1980, \apjs, 43, 469

\bibitem[{{Tkachenko} {et~al.}(2013{\natexlab{a}}){Tkachenko}, {Aerts},
  {Yakushechkin}, {Debosscher}, {Degroote}, {Bloemen}, {P{\'a}pics}, {de
  Vries}, {Lombaert}, {Hrudkova}, {Fr{\'e}mat}, {Raskin}, \& {Van
  Winckel}}]{Tkachenko2013}
{Tkachenko}, A., {Aerts}, C., {Yakushechkin}, A., {et~al.} 2013{\natexlab{a}},
  \aap, 556, A52

\bibitem[{{Tkachenko} {et~al.}(2013{\natexlab{b}}){Tkachenko}, {Lehmann},
  {Smalley}, \& {Uytterhoeven}}]{Tkachenko2013c}
{Tkachenko}, A., {Lehmann}, H., {Smalley}, B., \& {Uytterhoeven}, K.
  2013{\natexlab{b}}, \mnras, 431, 3685

\bibitem[{{Tkachenko} {et~al.}(2013{\natexlab{c}}){Tkachenko}, {Van Reeth},
  {Tsymbal}, {Aerts}, {Kochukhov}, \& {Debosscher}}]{Tkachenko2013b}
{Tkachenko}, A., {Van Reeth}, T., {Tsymbal}, V., {et~al.} 2013{\natexlab{c}},
  \aap, 560, A37

\bibitem[{{Townsend}(2003)}]{Townsend2003}
{Townsend}, R.~H.~D. 2003, \mnras, 343, 125

\bibitem[{{Townsend} \& {Teitler}(2013)}]{Townsend2013}
{Townsend}, R.~H.~D. \& {Teitler}, S.~A. 2013, \mnras, 435, 3406

\end{thebibliography}

\end{document}